\documentclass[% 
 reprint,
 amsmath,amssymb,
 aps,
]{revtex4-2}

\usepackage{graphicx}% Include figure files
\usepackage{tikz}
\usepackage{amsmath}
\usepackage{pgfplots}
\usepackage{subcaption}
\pgfplotsset{compat=1.18}
\usepackage{changepage}
\usepackage{float}
\usepackage{dcolumn}% Align table columns on decimal point
\usepackage{bm}% bold math
\begin{document}

\preprint{APS/123-QED}

\title{Dynamics of quantum measurement via electron transport in quantum dot systems: many-particle wavefunction approach}% Force line breaks with \\

\author{George Stavisskii}
\email{stavisski.gl@phystech.edu}
\author{Leonid Fedichkin}
\email{leonid@phystech.edu}
\affiliation{L.D. Landau Dept. of Theoretical Physics, Moscow Institute of Physics and Technology, Dolgoprudny, Russia\\%
             }

\date{\today}% It is always \today, today,

\begin{abstract}
To describe the measurement of a quantum system, one often incorporates the "surroundings" into the unitary evolution and then traces out the excess degrees of freedom associated with the measurement outcome, thereby obtaining irreversible dynamics. An alternative approach based on the many-particle wave function is presented, in addition to the widely used Keldysh, scattering matrix and TCL formalisms, built on the many-particle wave function, together with a generalization of the resulting equations to arbitrary quantum dot–point contact measurement systems.
\begin{description}
\item[Subject Areas]
Quantum transport, Quantum measurement, Quantum computation
\end{description}
\end{abstract}

\maketitle

%\tableofcontents

\section{\label{sec:level1}Introduction}
The growing interest in quantum technologies, particularly quantum evaluations, has led to the study of phenomena such as decoherence and relaxation in quantum systems which arise due to interactions with the surroundings and the measurement process. One of the most convenient tools for quantifying these aspects, as well as the overall dynamics of a quantum system, is the density operator. Using the formalism of density operators, one can relate the quantum properties of an arbitrary system to the classical output of a measurement device, in our case, a point contact. Quantum point contacts (QPCs) are widely used in the area to measure and analyze coherent and statistical properties of the quantum dot and tunneling current \cite{QPC_counting_first, QPC_FCS_2006} and thus present a convenient platform for the development of the theoretical formalism, which could, though, further be applied to arbitrary fermion baths. 

There are several conventional methods for describing irreversible dynamics, measurement processes, and relaxation in quantum systems, such as a qubit, as well as related non-equilibrium phenomena like electron transport. The first is the Keldysh formalism for open quantum systems. Although powerful, the Keldysh formalism often results in an overly complex treatment of measurement and transport dynamics, yielding general non-analytical solutions that require numerical analysis. In certain specific cases where simplifications are possible, a more straightforward theoretical approach may be advantageous. Another strong formalism for the theoretical analysis of electron transport in mesoscopic systems is the scattering matrix approach and full counting statistics, developed by Levitov and Lesovik \cite{Levitov_Lesovik_FCS_1996, Lesovik_Scattering_review}. This complex of methods can be quite attractive, since it allows one to study tunneling current starting directly from physical parameters, without Bardeen transfer Hamiltonian, and derive full statistics of the resulting transport up to any corresponding moments. However, one can be interested in time-dependent properties of current, mainly noise power spectrum, and inclusion of measurement by arbitrary device and it's effect on previously mentioned statistics. The scattering matrix approach does not allow it from the box and requires phenomenological "voltage probes" to introduce decoherence into transport and quantum dot evolution \cite{Coupled_Dots_2006, Pilgram_Path_decoherence_2003}. Thus, this apparatus is not always the most convenient for QPC-measurement analysis. Lastly, there is a variety of different approaches which focus on the Non-Markovian dynamics regimes of evolution in systems which consist of coupled quantum dots and fermionic leads \cite{TCL_formalism_paper, xue2015nonmarkoviancountingstatisticsquantum}. Since those formalisms such as the TCL projection operator method rely on the interaction picture expansion, they are perturbative in both time and interaction strength, which limits their application in problems of long-time behavior. Moreover, they often lead to overcomplicated expressions for resulting master-equations.

To address those issues, we formalize and use the many-particle wave function approach, originally developed in previous works by Gurvitz \cite{gurvitz2016wavefunctionapproachmasterequations}, and apply it to transport in quantum dot–point contact systems coupled to a qubit via Coulomb interaction. We further generalize the obtained Master equations and measurement statistics to arbitrary one-electron-transistor-like structures. 

\section{\label{sec:level1} Point contact as a framework for Many-body wavefunction approach}

Now, we turn to a simple example of a quantum dot–point contact system, which we will refer to as the "2-D quantum bottleneck," to present a formalized derivation of the many-particle wavefunction approach to the kinetics of electronic transport and subsequently generalize the results to more complex PC structures, which can show different non-trivial properties of the tunneling current and full counting statistics \cite{Coupled_Dots_2006, dynamical_double_blockade_2005, Spin_dynamical_blockade_CC}.

Hamiltonian of the corresponding system:

\begin{equation}
    H = H_{t} + \delta \Omega {a_{0}}^\dagger a_{0} H_{t} + H_{pc} + H_{q} 
\end{equation}

where:

\begin{equation}
\begin{gathered}
    H_{t} = \sum_{l} \Omega_{l\alpha}{a_{\alpha}}^\dagger a_{l} + \Omega_{l\beta}{a_{\beta}}^\dagger a_{l} + H.C \\
    + \sum_{r} \Omega_{r\alpha}{a_{\alpha}}^\dagger a_{r} + \Omega_{r\beta}{a_{\beta}}^\dagger a_{r} + H.C \\
\end{gathered}
\end{equation}

\begin{equation}
\begin{gathered}    
    H_{pc} =  \sum_{l} E_{l} a_{l}^{\dagger}a_{l} + \sum_{r} E_{r} a_{r}^{\dagger}a_{r}     
\end{gathered}
\end{equation}

\begin{equation}
\begin{gathered}
    H_{q} = E_{0} a_{0}^{\dagger} a_{0} + E_{1} a_{1}^{\dagger} a_{1} + \epsilon(a_{1}^{\dagger}a_{0} + a_{0}a_{1}^{\dagger})
\end{gathered}
\end{equation}

Our PC system graphical representation is shown in fig. 1.

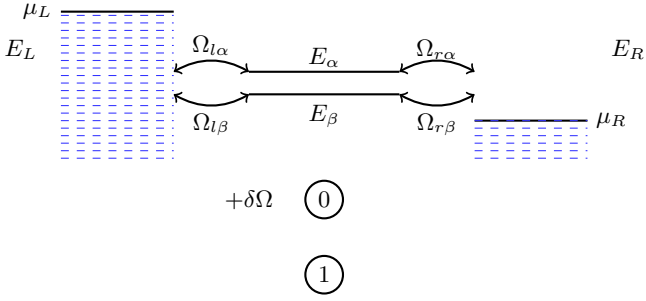
\begin{figure}
\begin{tikzpicture}[scale=1]
    % Energy band on the left (mu_L)
    \draw[thick] (-3.5, 2) -- (-2, 2);  % Fermi energy line on the left
    \node[left] at (-3.5, 2) {$\mu_L$};  % Label for Fermi energy level on the left

    % Energy band with dashed deep blue lines for left band
    \foreach \i in {0.05, 0.15, 0.25, 0.35, 0.45, 0.55, 0.65, 0.75, 0.85, 0.95, 1.05, 1.15, 1.25, 1.35, 1.45, 1.55, 1.65, 1.75, 1.85, 1.95} {
        \draw[dashed, blue!80] (-3.5, \i) -- (-2, \i);  % Dashed lines for left energy band in deep blue
    }

    % Energy band on the right (mu_R)
    \draw[thick] (2, 0.55) -- (3.5, 0.55);  % Fermi energy line on the right
    \node[right] at (3.5, 0.55) {$\mu_R$};  % Label for Fermi energy level on the right

    % Energy band with dashed deep blue lines for right band
    \foreach \i in {0.05, 0.15, 0.25, 0.35, 0.45, 0.55} {
        \draw[dashed, blue!80] (2, \i) -- (3.5, \i);  % Dashed lines for right energy band in deep blue
    }

    % Labels for energy bands on the sides
    \node[left] at (-3.7, 1.5) {$E_L$};  % Label for left band
    \node[right] at (3.7, 1.5) {$E_R$};  % Label for right band

    % Two energy levels in the middle, closer and lower in the band
    \draw[thick] (-1, 1.2) -- (1, 1.2);  % Alpha level, placed lower in the band

    \draw[thick] (-1, 0.9) -- (1, 0.9);  % Beta level, placed lower and closer to alpha

    % Labels for energy levels E_alpha and E_beta, adjusted for clarity
    \node[above] at (0, 1.15) {$E_\alpha$};  % Label for alpha energy level
    \node[above] at (0, 0.35) {$E_\beta$};   % Label for beta energy level

    % Curvilinear double-sided arrows from left band to energy levels with omega labels
    \draw[<->, thick, bend left=30] (-2.0, 1.2) to (-1.0, 1.2);  % Arrow for alpha level
    \node at (-1.5, 1.55) {$\Omega_{l\alpha}$};  % Omega label for alpha level arrow

    \draw[<->, thick, bend left=-30] (-2.0, 0.9) to (-1.0, 0.9);   % Arrow for beta level
    \node at (-1.5, 0.5) {$\Omega_{l\beta}$};   % Omega label for beta level arrow

    % Curvilinear double-sided arrows from right band to energy levels with omega labels
    \draw[<->, thick, bend right=30] (2.0, 1.2) to (1.0, 1.2);  % Arrow for alpha level
    \node at (1.5, 1.5) {$\Omega_{r\alpha}$};  % Omega label for alpha level arrow

    \draw[<->, thick, bend right=-30] (2.0, 0.9) to (1.0, 0.9);   % Arrow for beta level
    \node at (1.5, 0.5) {$\Omega_{r\beta}$};   % Omega label for beta level arrow

    % Two quantum dots for qubit representation below, positioned vertically and centered
    \draw[thick] (0, -0.5) circle (0.25);  % Left quantum dot for state 0
    \draw[thick] (0, -1.5) circle (0.25);  % Right quantum dot for state 1

    % Labels for quantum dot states (without ket notation)
    \node at (0, -0.5) {0};  % Label for the state 0
    \node at (-1, -0.5) {$+\delta\Omega$};  % Label for the state 0
    \node at (0, -1.5) {1};  % Label for the state 1

\end{tikzpicture}
\caption{PC-2D quantum bottleneck energy band diagrammatic representation.}
\end{figure}

Next, we will proceed with the many-body wavefunction formalism, as developed in the works of Gurvitz \cite{gurvitz2016wavefunctionapproachmasterequations, Gurvitz_1997}. First, we examine the case of $\delta \Omega = 0$, where the current in the point contact detector is independent of the qubit dynamics, and discuss in detail the derivation and solution of the PC equations. Afterwards, we apply analogous methods to derive the previously mentioned Master equations for the reduced density operator of the qubit system. Up to this point, we leave the discussion of decoherence due to relaxation to some bosonic spectra aside, though we will comment on how one can include it in the results in simple cases.

\section{\label{sec:level1} Point contact detector with 2-d "bottleneck"}

From fermionic nature of operators in system, it is clear that wave function of whole PC can be written as:

\begin{equation}
\begin{gathered}
    \lvert\psi\rangle = (b_{0} + \sum_{l, r} b_{lr} a_{r}^{\dagger}a_{l} + \sum_{l} b_{l\alpha} a_{\alpha}^{\dagger}a_{l} + \sum_{l} b_{l\beta} a_{\beta}^{\dagger}a_{l} \\
    + \sum_{l < l'} b_{ll'\alpha\beta} a_{\alpha}^{\dagger}a_{\beta}^{\dagger}a_{l}a_{l'} + ...)|0\rangle
\end{gathered}
\end{equation}

Now we are interested in isolation of all $b_{ll'...\alpha\beta rr'...}$ amplitudes in Shrodinger equation: $i\frac{d|\psi\rangle}{dt} = (H_{t} + H_{pc})|\psi\rangle$.

The main analytical instrument of this work, which we formalize and develop, is the many-body wavefunction method, first proposed by Gurvitz \cite{gurvitz2016wavefunctionapproachmasterequations}. It provides a very concise treatment of electron transport in discussed systems. Moreover, it allows an elegant inclusion of the measurement procedure via direct interaction of the electrons in leads and a considered discrete quantum system. Using such advantage, one can also rigorously treat transition between coherent and in-coherent tunneling regimes in PC's without phenomenological constructions. A rigorous derivation of the method is provided, including many additional details that have not been discussed previously. For example, we obtain full expressions for wavefunction amplitudes with arbitrary number of particles and discuss general simplification procedure.

Thus, we can categorize all the amplitudes in the system into three groups: no electrons in the 2-D bottleneck, one electron, and two electrons. We do not account for spin degrees of freedom in this discussion due to Coloumb blockade in the PC.

In the case of the "alpha" term, we clearly identify four different ways to obtain it by the action of the transition part of the Hamiltonian, which are graphically represented by the fig. 2. 

Derivation for any amplitude further will be treated in terms of such "hopping" diagrams, since in our Bardeen transfer Hamiltonian, interaction part, $H_{t}$, contains only pairs of creation/annihilation operators of the electrons in both leads. Each diagram results precisely in one of four possible contributions of interaction terms into the final equation, and provides one a convenient way to formally derive Shrodinger equation in a many-body representation. Same logic holds for any number of state inside considered PC.

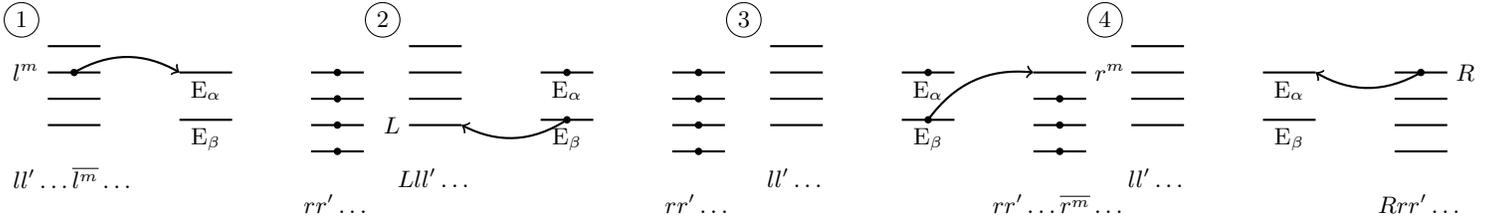
\begin{figure*}[t]
\hspace{-2cm}
\begin{subfigure}{0.2\textwidth}
    \centering
    \begin{tikzpicture}[scale=0.7]
        % Add a circle with the number 1 in the top left corner
        \node[circle, draw, inner sep=2pt] at (-3.5, 3) {1};  % Circle with number 1
    
        % Left side: four energy levels
        \foreach \y in {1, 1.5, 2, 2.5} {
            \draw[thick] (-3, \y) -- (-2, \y);  % Energy levels on the left
        }
    
        % Add an electron (dot) on the level in the left band from which the arrow points
        \fill (-2.5, 2) circle (2pt);  % Electron (dot) on the level at y = 2
    
        % Label for the level with the electron
        \node[left] at (-3, 2) {$l^m$};  % Label l^m for the level with the electron
    
        % Label for the entire left spectra below
        \node at (-2.5, 0) {$ll' \dots \overline{l^m}\dots$};
    
        % Middle: two energy levels, swapped alpha and beta
        \foreach \y in {1.1, 2.0} {
            \draw[thick] (-0.5, \y) -- (0.5, \y);  % Energy levels in the middle
        }
    
        % Labels for middle energy levels (swapped alpha and beta)
        \node[below] at (0, 1.1) {E$_\beta$};
        \node[below] at (0, 2.0) {E$_\alpha$};
    
        % Right side: four energy levels, moved lower and occupied by electrons
        \foreach \y in {0.5, 1, 1.5, 2.0} {
            \draw[thick] (2, \y) -- (3, \y);  % Energy levels on the right
            \fill (2.5, \y) circle (2pt);  % Electron (dot) on each energy level
        }
    
        % Label for the entire right spectra below
        \node at (2.5, -0.5) {$rr' \dots$};
    
        % Curvilinear arrow from left band to alpha level
        \draw[->, thick, bend left=30] (-2.5, 2) to (-0.5, 2.0);  % Arrow from left band to alpha level
    
    \end{tikzpicture}
\end{subfigure}
\hspace{1cm}
\begin{subfigure}{0.2\textwidth}
\centering
\begin{tikzpicture}[scale=0.7]
    % Add a circle with the number 1 in the top left corner
    \node[circle, draw, inner sep=2pt] at (-3.5, 3) {2};  % Circle with number 2

    % Left side: four energy levels
    \foreach \y in {1, 1.5, 2, 2.5} {
        \draw[thick] (-3, \y) -- (-2, \y);  % Energy levels on the left
    }

    % Label for the entire left spectra below
    \node at (-2.5, 0) {$Lll'\dots$};

    % Middle: two energy levels, swapped alpha and beta
    \foreach \y in {1.1, 2.0} {
        \draw[thick] (-0.5, \y) -- (0.5, \y);  % Energy levels in the middle
    }

    % Labels for middle energy levels (swapped alpha and beta)
    \node[below] at (0, 1.1) {E$_\beta$};
    \node[below] at (0, 2.0) {E$_\alpha$};

    % Add electrons (dots) on the middle energy levels
    \fill (0, 1.1) circle (2pt);  % Electron on E_beta level
    \fill (0, 2.0) circle (2pt);  % Electron on E_alpha level

    % Right side: four energy levels, moved lower and occupied by electrons
    \foreach \y in {0.5, 1, 1.5, 2.0} {
        \draw[thick] (2, \y) -- (3, \y);  % Energy levels on the right
        \fill (2.5, \y) circle (2pt);  % Electron (dot) on each energy level
    }

    % Label for the entire right spectra below
    \node at (2.5, -0.5) {$rr' \dots$};

    % Curvilinear arrow from beta level pointing into the left band
    \draw[->, thick, bend left=30] (0, 1.1) to (-2.0, 1);  % Arrow from beta level to left band

    \node[left] at (-3, 1) {$L$};

\end{tikzpicture}
\end{subfigure}
\hspace{1cm}
\begin{subfigure}{0.2\textwidth}
\centering
\begin{tikzpicture}[scale=0.7]
    % Add a circle with the number 3 in the top left corner
    \node[circle, draw, inner sep=2pt] at (-3.5, 3) {3};  % Circle with number 3

    % Left side: four energy levels
    \foreach \y in {1, 1.5, 2, 2.5} {
        \draw[thick] (-3, \y) -- (-2, \y);  % Energy levels on the left
    }

    % Label for the entire left spectra below
    \node at (-2.5, 0) {$ll'\dots$};

    % Middle: two energy levels, swapped alpha and beta
    \foreach \y in {1.1, 2.0} {
        \draw[thick] (-0.5, \y) -- (0.5, \y);  % Energy levels in the middle
    }

    % Labels for middle energy levels (swapped alpha and beta)
    \node[below] at (0, 1.1) {E$_\beta$};
    \node[below] at (0, 2.0) {E$_\alpha$};

    % Add electrons (dots) on the middle energy levels
    \fill (0, 1.1) circle (2pt);  % Electron on E_beta level
    \fill (0, 2.0) circle (2pt);  % Electron on E_alpha level

    % Right side: four energy levels, moved lower and occupied by electrons
    \foreach \y in {0.5, 1, 1.5} {
        \draw[thick] (2, \y) -- (3, \y);  % Energy levels on the right
        \fill (2.5, \y) circle (2pt);  % Electron (dot) on each energy level
    }
    
    % Draw the upper energy level without an electron and add label r^m
    \draw[thick] (2, 2.0) -- (3, 2.0);  % Energy level on the right
    \node[right] at (3, 2.0) {$r^m$};  % Label r^m for the upper right energy level

    % Label for the entire right spectra below
    \node at (2.5, -0.5) {$rr'\dots \overline{r^{m}} \dots$};

    % Arrow from beta level to the upper level in the right band
    \draw[->, thick, bend left=30] (0, 1.1) to (2.0, 2.0);  % Arrow from beta level to upper right band level

\end{tikzpicture}
\end{subfigure}
\hspace{1cm}
\begin{subfigure}{0.2\textwidth}
\centering
\begin{tikzpicture}[scale=0.7]
    % Add a circle with the number 3 in the top left corner
    \node[circle, draw, inner sep=2pt] at (-3.5, 3) {4};  % Circle with number 4

    % Left side: four energy levels
    \foreach \y in {1, 1.5, 2, 2.5} {
        \draw[thick] (-3, \y) -- (-2, \y);  % Energy levels on the left
    }

    % Label for the entire left spectra below
    \node at (-2.5, 0) {$ll'\dots$};

    % Middle: two energy levels, swapped alpha and beta
    \foreach \y in {1.1, 2.0} {
        \draw[thick] (-0.5, \y) -- (0.5, \y);  % Energy levels in the middle
    }

    % Labels for middle energy levels (swapped alpha and beta)
    \node[below] at (0, 1.1) {E$_\beta$};
    \node[below] at (0, 2.0) {E$_\alpha$};

    % Right side: four energy levels, with an electron on the upper level
    \foreach \y in {0.5, 1, 1.5} {
        \draw[thick] (2, \y) -- (3, \y);  % Energy levels on the right
    }
    
    % Add electron (dot) on the upper right energy level
    \fill (2.5, 2.0) circle (2pt);  % Electron on the upper level

    % Draw the upper energy level without an electron and add label r^m
    \draw[thick] (2, 2.0) -- (3, 2.0);  % Energy level on the right
    \node[right] at (3.0, 2.0) {$R$};  % Label r^m for the upper right energy level

    % Label for the entire right spectra below
    \node at (2.5, -0.5) {$Rrr' \dots$};

    % Rearranged arrow from the upper right level to the alpha level in the middle
    \draw[->, thick, bend left=30] (2.5, 2.0) to (0.5, 2.0);  % Arrow from upper right level to alpha level

\end{tikzpicture}
\end{subfigure}
\caption{Terms in transition Hamiltonian, which give contribution to the "alpha" term. Indexes denote many-body wavefunction argument of the contributing term.}
\end{figure*}

After adding the trivial self-energy part in the equation for the desired amplitude, we get:

\vspace{-0.5cm}
\begin{equation}
\begin{gathered}
    i \dot{b_{ll'\dots\alpha rr' \dots}} = \sum_{m = 1}^{n} (-1)^{n + m} \Omega_{l^{m}\alpha} b_{ll' \dots \overline{l^{m}} \dots rr'} \\
    + (-1)^{n - 1} \sum_{L} \Omega_{L\beta} b_{Lll'\dots \alpha \beta rr' \dots} + \sum_{m = 1}^{n - 1} (-1)^{m - 1} \Omega_{r^{m}\beta} b_{ll' \dots \alpha \beta rr'\dots \overline{r^{m}} \dots} \\
     + \sum_{R} \Omega_{R \alpha} b_{ll'\dots Rrr' \dots} + (E_{\alpha} + \sum_{r, r' \dots} E_{R} - \sum_{l, l' \dots} E_{L})b_{ll'\dots \alpha 
     rr' \dots}
\end{gathered}
\end{equation}

\vspace{-0cm}

Signs appear in expressions due to fermionic anticommutation relation $\{a_{l}, a_{r}^{\dagger}\} = \delta_{lr}$, $n$ is the number of holes, and $m$ runs over collector and emitter indices. Graphical representation of $"\alpha \beta"$ term can be seen in fig. 3.

\vspace{2cm}

\begin{figure}[H]
\begin{tikzpicture}[scale=1]
    % Add a circle with the number 1 in the top left corner
    \node[circle, draw, inner sep=2pt] at (-3.5, 3) {1};  % Circle with number 1

    % Left side: four energy levels
    \foreach \y in {1, 1.5, 2, 2.5} {
        \draw[thick] (-3, \y) -- (-2, \y);  % Energy levels on the left
    }

    % Add an electron (dot) on the level in the left band from which the arrow points
    \fill (-2.5, 2) circle (2pt);  % Electron (dot) on the level at y = 2

    % Label for the level with the electron
    \node[left] at (-3, 2) {$l^m$};  % Label l^m for the level with the electron

    % Label for the entire left spectra below
    \node at (-2.5, 0) {$ll' \dots \overline{l^m}\dots$};

    % Middle: two energy levels, swapped alpha and beta
    \foreach \y in {1.1, 2.0} {
        \draw[thick] (-0.5, \y) -- (0.5, \y);  % Energy levels in the middle
    }

    \fill (0, 1.1) circle (2pt);

    % Labels for middle energy levels (swapped alpha and beta)
    \node[below] at (0, 1.1) {E$_\beta$};
    \node[below] at (0, 2.0) {E$_\alpha$};

    % Right side: four energy levels, moved lower and occupied by electrons
    \foreach \y in {0.5, 1, 1.5, 2.0} {
        \draw[thick] (2, \y) -- (3, \y);  % Energy levels on the right
        \fill (2.5, \y) circle (2pt);  % Electron (dot) on each energy level
    }

    % Label for the entire right spectra below
    \node at (2.5, -0.5) {$rr' \dots$};

    % Curvilinear arrow from left band to alpha level
    \draw[->, thick, bend left=30] (-2.5, 2) to (-0.5, 2.0);  % Arrow from left band to alpha level

\end{tikzpicture}

\begin{tikzpicture}[scale=1]
    % Add a circle with the number 1 in the top left corner
    \node[circle, draw, inner sep=2pt] at (-3.5, 3) {2};  % Circle with number 2

    % Left side: four energy levels
    \foreach \y in {1, 1.5, 2, 2.5} {
        \draw[thick] (-3, \y) -- (-2, \y);  % Energy levels on the left
    }

    % Label for the entire left spectra below
    \node at (-2.5, 0) {$ll' \dots \overline{l^m}\dots$};

    % Label for the level with the electron
    \node[left] at (-3, 1) {$l^{m}$};  % Label l^m for the level with the electron

    % Middle: two energy levels, swapped alpha and beta
    \foreach \y in {1.1, 2.0} {
        \draw[thick] (-0.5, \y) -- (0.5, \y);  % Energy levels in the middle
    }

    \fill (0, 2.0) circle (2pt);

    \fill (-2.5, 1.0) circle (2pt);

    % Labels for middle energy levels (swapped alpha and beta)
    \node[below] at (0, 1.1) {E$_\beta$};
    \node[below] at (0, 2.0) {E$_\alpha$};

    % Right side: four energy levels, moved lower and occupied by electrons
    \foreach \y in {0.5, 1, 1.5, 2.0} {
        \draw[thick] (2, \y) -- (3, \y);  % Energy levels on the right
        \fill (2.5, \y) circle (2pt);  % Electron (dot) on each energy level
    }

    % Label for the entire right spectra below
    \node at (2.5, -0.5) {$rr' \dots$};

    % Curvilinear arrow from left band to alpha level
    \draw[->, thick, bend left=30] (-2.5, 1) to (-0.5, 1.1);  % Arrow from left band to alpha level

\end{tikzpicture}

\begin{tikzpicture}[scale=1]
    % Add a circle with the number 1 in the top left corner
    \node[circle, draw, inner sep=2pt] at (-3.5, 3) {3};  % Circle with number 3

    % Left side: four energy levels
    \foreach \y in {1, 1.5, 2, 2.5} {
        \draw[thick] (-3, \y) -- (-2, \y);  % Energy levels on the left
    }

    % Label for the level with the electron
    \node[left] at (-3, 1) {$l^m$};  % Label l^m for the level with the electron

    % Label for the entire left spectra below
    \node at (-2.5, 0) {$ll'\dots$};

    % Middle: two energy levels, swapped alpha and beta
    \foreach \y in {1.1, 2.0} {
        \draw[thick] (-0.5, \y) -- (0.5, \y);  % Energy levels in the middle
    }

    \fill (0, 1.1) circle (2pt);

    % Label for the level with the electron
    \node[right] at (3, 2) {$R$};  % Label l^m for the level with the electron

    %\fill (-2.5, 1.0) circle (2pt);

    % Labels for middle energy levels (swapped alpha and beta)
    \node[below] at (0, 1.1) {E$_\beta$};
    \node[below] at (0, 2.0) {E$_\alpha$};

    % Right side: four energy levels, moved lower and occupied by electrons
    \foreach \y in {0.5, 1, 1.5, 2.0} {
        \draw[thick] (2, \y) -- (3, \y);  % Energy levels on the right
        \fill (2.5, \y) circle (2pt);  % Electron (dot) on each energy level
    }

    % Label for the entire right spectra below
    \node at (2.5, -0.5) {$Rrr' \dots$};

    % Curvilinear arrow from left band to alpha level
    \draw[->, thick, bend left=30] (2.5, 2) to (0.5, 2);  % Arrow from left band to alpha level

\end{tikzpicture}

\begin{tikzpicture}[scale=1]
    % Add a circle with the number 1 in the top left corner
    \node[circle, draw, inner sep=2pt] at (-3.5, 3) {4};  % Circle with number 4

    % Left side: four energy levels
    \foreach \y in {1, 1.5, 2, 2.5} {
        \draw[thick] (-3, \y) -- (-2, \y);  % Energy levels on the left
    }

    % Label for the level with the electron
    \node[left] at (-3, 1) {$l^m$};  % Label l^m for the level with the electron

    % Label for the entire left spectra below
    \node at (-2.5, 0) {$ll'\dots$};

    % Middle: two energy levels, swapped alpha and beta
    \foreach \y in {1.1, 2.0} {
        \draw[thick] (-0.5, \y) -- (0.5, \y);  % Energy levels in the middle
    }

    \fill (0, 2) circle (2pt);

    % Label for the level with the electron
    \node[right] at (3, 2) {$R$};  % Label l^m for the level with the electron

    %\fill (-2.5, 1.0) circle (2pt);

    % Labels for middle energy levels (swapped alpha and beta)
    \node[below] at (0, 1.1) {E$_\beta$};
    \node[below] at (0, 2.0) {E$_\alpha$};

    % Right side: four energy levels, moved lower and occupied by electrons
    \foreach \y in {0.5, 1, 1.5, 2.0} {
        \draw[thick] (2, \y) -- (3, \y);  % Energy levels on the right
        \fill (2.5, \y) circle (2pt);  % Electron (dot) on each energy level
    }

    % Label for the entire right spectra below
    \node at (2.5, -0.5) {$Rrr' \dots$};

    % Curvilinear arrow from left band to alpha level
    \draw[->, thick, bend left=30] (2.5, 2) to (0.5, 1.1);  % Arrow from left band to alpha level
\end{tikzpicture}
\caption{Terms in transition Hamiltonian, which give contribution to the $\alpha \beta$ term. Indexes denote many-body wavefunction argument of the contributing term.}
\end{figure}

And thus, we get:

\[
    i\dot{b_{ll'\dots \alpha \beta rr'}} = \sum_{m = 1}^{n} (-1)^{n + m} \Omega_{l^{m}\alpha} b_{ll'\dots \overline{l^{m}} \dots \beta rr'}
\]

\[
    - \sum_{m = 1}^{n} (-1)^{n + m} \Omega_{l^{m}\beta} b_{ll'\dots \overline{l^{m}} \dots \alpha rr'} - \sum_{R} \Omega_{R\alpha} b_{ll'\dots \beta Rrr'\dots}
\]

\begin{equation}
    + \sum_{R} \Omega_{R\beta} b_{ll'\dots \alpha Rrr'\dots} + (E_{\alpha} + E_{\beta} + \sum_{r, r' \dots} E_{R} - \sum_{l, l \dots} E_{L}) b_{ll'\dots \alpha \beta rr' \dots}
\end{equation}

And lastly, term "0" without electrons in 2-d bottleneck can be seen in fig. 4.

\begin{figure}
\begin{tikzpicture}[scale=1]
    % Add a circle with the number 1 in the top left corner
    \node[circle, draw, inner sep=2pt] at (-3.5, 3) {1};  % Circle with number 1

    % Left side: four energy levels
    \foreach \y in {1, 1.5, 2, 2.5} {
        \draw[thick] (-3, \y) -- (-2, \y);  % Energy levels on the left
    }

    % Label for the level with the electron
    \node[left] at (-3, 2.5) {$L$};  % Label l^m for the level with the electron

    % Label for the entire left spectra below
    \node at (-2.5, 0) {$Lll'\dots$};

    % Middle: two energy levels, swapped alpha and beta
    \foreach \y in {1.1, 2.0} {
        \draw[thick] (-0.5, \y) -- (0.5, \y);  % Energy levels in the middle
    }

    \fill (0, 2) circle (2pt);

    %\fill (-2.5, 1.0) circle (2pt);

    % Labels for middle energy levels (swapped alpha and beta)
    \node[below] at (0, 1.1) {E$_\beta$};
    \node[below] at (0, 2.0) {E$_\alpha$};

    % Right side: four energy levels, moved lower and occupied by electrons
    \foreach \y in {0.5, 1, 1.5, 2.0} {
        \draw[thick] (2, \y) -- (3, \y);  % Energy levels on the right
        \fill (2.5, \y) circle (2pt);  % Electron (dot) on each energy level
    }

    % Label for the entire right spectra below
    \node at (2.5, -0.5) {$rr' \dots$};

    % Curvilinear arrow from left band to alpha level
    \draw[->, thick, bend left=-30] (0, 2) to (-2.0, 2.5);  % Arrow from left band to alpha level

\end{tikzpicture}

\begin{tikzpicture}[scale=1]
    % Add a circle with the number 1 in the top left corner
    \node[circle, draw, inner sep=2pt] at (-3.5, 3) {2};  % Circle with number 2

    % Left side: four energy levels
    \foreach \y in {1, 1.5, 2, 2.5} {
        \draw[thick] (-3, \y) -- (-2, \y);  % Energy levels on the left
    }

    % Label for the level with the electron
    \node[left] at (-3, 2.5) {$L$};  % Label l^m for the level with the electron

    % Label for the entire left spectra below
    \node at (-2.5, 0) {$Lll'\dots$};

    % Middle: two energy levels, swapped alpha and beta
    \foreach \y in {1.1, 2.0} {
        \draw[thick] (-0.5, \y) -- (0.5, \y);  % Energy levels in the middle
    }

    \fill (0, 1.1) circle (2pt);

    %\fill (-2.5, 1.0) circle (2pt);

    % Labels for middle energy levels (swapped alpha and beta)
    \node[below] at (0, 1.1) {E$_\beta$};
    \node[below] at (0, 2.0) {E$_\alpha$};

    % Right side: four energy levels, moved lower and occupied by electrons
    \foreach \y in {0.5, 1, 1.5, 2.0} {
        \draw[thick] (2, \y) -- (3, \y);  % Energy levels on the right
        \fill (2.5, \y) circle (2pt);  % Electron (dot) on each energy level
    }

    % Label for the entire right spectra below
    \node at (2.5, -0.5) {$rr' \dots$};

    % Curvilinear arrow from left band to alpha level
    \draw[->, thick, bend left=-30] (0, 1.1) to (-2.0, 2.5);  % Arrow from left band to alpha level

\end{tikzpicture}

\begin{tikzpicture}[scale=1]
    % Add a circle with the number 1 in the top left corner
    \node[circle, draw, inner sep=2pt] at (-3.5, 3) {3};  % Circle with number 3

    % Left side: four energy levels
    \foreach \y in {1, 1.5, 2, 2.5} {
        \draw[thick] (-3, \y) -- (-2, \y);  % Energy levels on the left
    }

    % Label for the level with the electron
    \node[right] at (3, 2) {$r^{m}$};  % Label l^m for the level with the electron

    % Label for the entire left spectra below
    \node at (-2.5, 0) {$ll'\dots$};

    % Middle: two energy levels, swapped alpha and beta
    \foreach \y in {1.1, 2.0} {
        \draw[thick] (-0.5, \y) -- (0.5, \y);  % Energy levels in the middle
    }

    \fill (0, 2) circle (2pt);

    %\fill (-2.5, 1.0) circle (2pt);

    % Labels for middle energy levels (swapped alpha and beta)
    \node[below] at (0, 1.1) {E$_\beta$};
    \node[below] at (0, 2.0) {E$_\alpha$};

    % Right side: four energy levels, moved lower and occupied by electrons
    \foreach \y in {0.5, 1, 1.5} {
        \draw[thick] (2, \y) -- (3, \y);  % Energy levels on the right
        \fill (2.5, \y) circle (2pt);  % Electron (dot) on each energy level
    }

    \draw[thick] (2, 2) -- (3, 2);

    % Label for the entire right spectra below
    \node at (2.5, -0.5) {$rr' \dots \overline{r^{m}}\dots$};

    % Curvilinear arrow from left band to alpha level
    \draw[->, thick, bend left=30] (0, 2) to (2.0, 2.0);  % Arrow from left band to alpha level

\end{tikzpicture}

\begin{tikzpicture}[scale=1]
    % Add a circle with the number 1 in the top left corner
    \node[circle, draw, inner sep=2pt] at (-3.5, 3) {4};  % Circle with number 4

    % Left side: four energy levels
    \foreach \y in {1, 1.5, 2, 2.5} {
        \draw[thick] (-3, \y) -- (-2, \y);  % Energy levels on the left
    }

    % Label for the level with the electron
    \node[right] at (3, 2) {$r^{m}$};  % Label l^m for the level with the electron

    % Label for the entire left spectra below
    \node at (-2.5, 0) {$ll'\dots$};

    % Middle: two energy levels, swapped alpha and beta
    \foreach \y in {1.1, 2.0} {
        \draw[thick] (-0.5, \y) -- (0.5, \y);  % Energy levels in the middle
    }

    \fill (0, 1.1) circle (2pt);

    %\fill (-2.5, 1.0) circle (2pt);

    % Labels for middle energy levels (swapped alpha and beta)
    \node[below] at (0, 1.1) {E$_\beta$};
    \node[below] at (0, 2.0) {E$_\alpha$};

    % Right side: four energy levels, moved lower and occupied by electrons
    \foreach \y in {0.5, 1, 1.5} {
        \draw[thick] (2, \y) -- (3, \y);  % Energy levels on the right
        \fill (2.5, \y) circle (2pt);  % Electron (dot) on each energy level
    }

    \draw[thick] (2, 2) -- (3, 2);

    % Label for the entire right spectra below
    \node at (2.5, -0.5) {$rr' \dots \overline{r^{m}}\dots$};

    % Curvilinear arrow from left band to alpha level
    \draw[->, thick, bend left=30] (0, 1.1) to (2.0, 2.0);  % Arrow from left band to alpha level

\end{tikzpicture}
\caption{Terms in transition Hamiltonian, which give contribution to the $0$ term. Indexes denote many-body wavefunction argument of the contributing term.}
\end{figure}

We will finally obtain:

\[
    i\dot{b_{ll'\dots rr'\dots}} = \sum_{L} (-1)^{n} \Omega_{L\alpha} b_{Lll'\dots \alpha rr' \dots} + \sum_{L} (-1)^{n} \Omega_{L\beta} b_{Lll'\dots \beta rr'\dots} 
\]

\[
    + \sum_{m = 1}^{n} (-1)^{m - 1} \Omega_{r^{m}\alpha}b_{ll'\dots \alpha rr' \dots \overline{r^{m}} \dots} + \sum_{m = 1}^{n}\Omega_{r^{m}\beta}b_{ll'\dots \beta rr'\dots \overline{r^{m}}\dots} 
\]

\begin{equation}
    \label{eq: shred_many_body}
    + (\sum_{r, r'\dots}E_{R} - \sum_{l, l' \dots} E_{L})b_{ll'\dots rr'\dots}
\end{equation}

Further, we will work with Laplace transform of those equations for imaginary contour:

\begin{equation}
    b(t) \rightarrow b(E) = \lim_{\delta \rightarrow 0} \int_{0}^{\infty} b(t)e^{(E + i\delta)it} dt
\end{equation}

Using basic integration by parts, one can state that the effect this transformation will have on the equations is the replacement of $i$ in front of each amplitude with $(E + i\delta)$ and introduction of $t = 0$ value of amplitude $b(0)$, which does not have any effect on our derivations.

Now, we need to simplify the obtained equations. We can split each right-hand side of the equations \eqref{eq: shred_many_body} into a "Spectral" part, which contains a summation over all states of the collector or emitter, and a "Discrete" part, which contains a summation over only a finite number of states. Next, we can insert the obtained equations for other terms into the "Spectral" parts. It can be seen that, since the quantum states at both the collector and emitter are very dense, we can replace the summation with an integration. By expressing the terms as sums using the equations obtained earlier, one can observe that terms of the form $(E + E_{L} + E_{l} + E_{l'} + \dots - E_{r} - E_{r'} - \dots)$ or $(E + E_{l} + E_{l'} + \dots - E_{R} - E_{r} - E_{r'} - \dots)$ will appear in the denominator. At this point, we can notice that all terms, except for one, will again contain an index over which we integrate. We can prove that all such terms, in the limit of high voltage, will vanish after integration. Indeed, once again expressing these terms using our core equations, they will end up looking like a sum of terms that will be of the form:

\begin{equation}
    ~ \int_{-\infty}^{\infty}dE_{L}\frac{C}{(E + i\delta + E_{L} + \dots) (E + i\delta + E_{L} + \dots)\dots}
\end{equation}

Here, we bring back the imaginary "renorm" part of the Laplace transform. It's easy to see that this integral can be treated using the complex technique of integration via the residue theorem. One can observe that all poles of the integrand lie below the real line, pole structrure is shown on fig. 5.

\begin{figure}
\hspace{0.5cm}
\begin{tikzpicture}[scale=1.5]
    % Draw the axes
    \draw[->] (-2, 0) -- (2, 0) node[below] {$\operatorname{Re}(E)$}; % Real axis
    \draw[->] (0, -2) -- (0, 2.5) node[left] {$\operatorname{Im}(E)$}; % Imaginary axis
    
    % Draw the semicircular contour in the upper half-plane with arrows for path of integration
    \draw[thick, ->] (1.5, 0) arc[start angle=0, end angle=45, radius=1.5];
    \draw[thick] (1.065, 1.065) arc[start angle=45, end angle=135, radius=1.5];
    \draw[thick, ->] (-1.055, 1.065) arc[start angle=135, end angle=180, radius=1.5];
    \draw[thick, ->] (0, 1.5) arc[start angle=90, end angle=135, radius=1.5];
    
    % Draw arrows along the real axis part of contour
    \draw[thick, ->] (-1.5, 0) -- (-0.0, 0); % Left half
    \draw[thick, ->] (0.0, 0) -- (1.5, 0); % Right half

    % Label the contour
    \node at (1.4, 1.0) {$\gamma_{R}$};

    \node at (0.5, 0.1) {$\gamma_{I}$};

    % Draw poles in the lower half-plane at the same vertical level
    \fill[black] (-1, -1) circle (1pt) node[left] {$z_1$}; % Pole 1
    \fill[black] (1, -1) circle (1pt) node[right] {$z_2$};  % Pole 2

    % Add labels for points on the contour
    \node at (1.5, 0) [below] {$E_{\lambda}$};
    \node at (-1.5, 0) [below] {$-E_{\lambda}$};

    \node at (0.2, -0.9) {$-i\delta$};

    \draw[thick, dashed] (-1,-1) -- (1,-1);

    % Optional: add a label at the origin
    \node at (0, 0) [below right] {$0$};
\end{tikzpicture}
\caption{Residue theorem explanation. $\gamma_{R}$ is the half-circle contour, $E_{\lambda}$ is the bias cutoff, $\gamma_{I}$ is the actual integration contour, $z_{1}$ and $z_{2}$ are poles of simplified integrand in (10).}
\end{figure}

So, from the structure of the integrand, we clearly see that $\int_{\gamma_{I} + \gamma_{R}} F(E)dE = 0$, since all poles are outside the contour for any value of $\delta > 0$. Additionally, since the power of the denominator is always $\geq 2$, $|\int_{\gamma_{R}} F(E)dE| \sim \frac{1}{E_{\lambda}}$, so in the limit of high voltage, we can extend the contour to infinity, and thus, $\int_{\gamma_{R}} F(E)dE \rightarrow 0$. Therefore, we conclude that $\int_{\gamma_{I}} F(E)dE \rightarrow 0$. We will use this important feature of such spectra-integrals extensively. \\

Finally, we focus on the only remaining term in the sum. In all cases, it's general form is:

\begin{equation}
    \int dE_{L, R}\frac{\rho_{L, R}\Omega_{(L, R)(\alpha, \beta)} \Omega_{(L, R)(\alpha, \beta)} \mathbf{b}(E)}{(E + i\delta + (-1)^{L, R}E_{L, R} + \dots)}
\end{equation}

In this case, in the final limit $\delta \rightarrow 0$, we can use the Sokhotski–Plemelj theorem. The result will be split into two parts: real and imaginary, principal and singular in context of the used theorem. In the limit of weak dependence of $\Omega$ and $\rho$ on energy levels (and $\Omega$ symmetry), we can neglect the principal parts in such integrals. Later, we will see that they vanish even without these considerations. The interesting part, the singular one, will yield:

\begin{equation}
    -i\pi\rho_{L, R}\Omega_{(L, R)(\alpha, \beta)}\Omega_{(L, R)(\alpha, \beta)}\mathbf{b(E)} = -i\frac{\Gamma_{(L, R)(\alpha, \beta)}}{2}\mathbf{b(E)}
\end{equation}

where we introduced the classical $\Gamma$ notation for energy widths relative to tunneling to the left and right bands of different energy levels in the middle 2-D "bottleneck".

After the process of "reducing" the spectral parts of the obtained equations, as described above, we get a simplified version of the previous equations:

"0" term:

\[
    (E + E_{l} + E_{l'} + \dots - E_{r} - E_{r'} - \dots + \frac{i\Gamma_{L\alpha}}{2} + \frac{i\Gamma_{L\beta}}{2})b_{ll'\dots rr'\dots} = 
\]

\begin{equation}
      \sum_{m = 1}^{n} (-1)^{m - 1} \Omega_{r^{m}\alpha}b_{ll'\dots \alpha rr' \dots \overline{r^{m}} \dots} + \sum_{m = 1}^{n}\Omega_{r^{m}\beta}b_{ll'\dots \beta rr'\dots \overline{r^{m}}\dots}
\end{equation}

$"\alpha\beta"$ term:

\[
    (E - E_{\alpha} - E_{\beta} + E_{l} + \dots - E_{r}+ \dots + \frac{i\Gamma_{R\alpha}}{2} + \frac{i\Gamma_{R\beta}}{2})b_{ll'\dots\alpha\beta rr'\dots} = 
\] 

\begin{equation}
    \hspace{-0.7cm}
    \sum_{m = 1}^{n} (-1)^{n + m} \Omega_{l^{m}\alpha} b_{ll'\dots \overline{l^{m}} \dots \beta rr'\dots} - \sum_{m = 1}^{n} (-1)^{n + m} \Omega_{l^{m}\beta} b_{ll'\dots \overline{l^{m}} \dots \alpha rr'\dots}
\end{equation}

and lastly, $\alpha$ and $\beta$ terms:

\[
    \hspace{-0.5cm}
    (E - E_{\alpha} + E_{l} + E_{l'} + \dots - E_{r} - E_{r'} + \dots + \frac{i\Gamma_{L\beta}}{2} + \frac{i\Gamma_{R\alpha}}{2})b_{ll'\dots\alpha rr'\dots} = 
\]

\[
    \frac{i\Delta\Gamma}{2}b_{ll'\dots\beta rr'\dots} + \sum_{m = 1}^{n} (-1)^{n + m} \Omega_{l^{m}\alpha} b_{ll' \dots \overline{l^{m}} \dots rr'\dots}
\]

\begin{equation}
    \label{eq: simplified_shred_many_body}
    + \sum_{m = 1}^{n - 1} (-1)^{m - 1} \Omega_{r^{m}\beta} b_{ll' \dots \alpha \beta rr'\dots \overline{r^{m}} \dots}
\end{equation}

\[
    (E - E_{\beta} + E_{l} + \dots - E_{r} - \dots + \frac{i\Gamma_{L\alpha}}{2} + \frac{i\Gamma_{R\beta}}{2})b_{ll'\dots\beta rr'\dots} = 
\]

\[
    \frac{i\Delta\Gamma}{2}b_{ll'\dots\alpha rr'\dots} + \sum_{m = 1}^{n} (-1)^{n + m} \Omega_{l^{m}\beta} b_{ll' \dots \overline{l^{m}} \dots rr'\dots}
\]

\begin{equation}
    - \sum_{m = 1}^{n - 1} (-1)^{m - 1} \Omega_{r^{m}\alpha} b_{ll' \dots \alpha \beta rr'\dots \overline{r^{m}} \dots}
\end{equation}

where $\Delta\Gamma = \Gamma_{L\alpha\beta} - \Gamma_{R\alpha\beta}$, which represents the coherent "tunneling" of an electron in the "bottleneck" from one level to another through the mechanism of current. We must note that for a large enough energy gap between these levels, this term vanishes ("independent current regime").

Now, we are ready to derive the equations for $\sigma^{n}$ reduced density operators of the middle "bottleneck", where $n$ stands for the number of electrons at the collector.

We obtain $\sigma^{n}$ by tracing out the collector and emitter states:

\begin{widetext}

\begin{equation}
\sigma_{(\alpha, \beta)(\alpha, \beta)}^{n} =
    \sum_{l < l' < \dots, r < r' \dots}\frac{1}{4\pi^{2}}\int b_{ll'\dots(\alpha, \beta) \dots rr'\dots}(E)b_{ll'\dots(\alpha, \beta) \dots rr'\dots}^{*}(E')e^{(i(E' - E)t)} dE dE'
\end{equation}

\end{widetext}

\normalsize

Next, for any equation \eqref{eq: simplified_shred_many_body} above, we multiply it by $b^{*}(E')$ and subtract from it the equation for $b^{*}(E')$ multiplied by $b(E)$. We will show an example for the $\sigma_{00}$ element:

\[
    \mathcal{L}^{-1}[(E - E' + i\Gamma_{L\alpha} + i\Gamma_{L\beta})b_{ll'\dots rr'\dots}(E)b_{ll'\dots rr'\dots}^{*}(E')] = 
\]

\[
    \hspace{-0.7cm}
    \mathcal{L}^{-1}[i * (\sum_{m = 1}^{n} (-1)^{m - 1} 2 \mathbf{Im}(\Omega_{r^{m}\alpha}b_{ll'\dots \alpha rr' \dots \overline{r^{m}} \dots}(E)b_{ll'\dots rr'\dots}^{*}(E')) + 
\]

\begin{equation}
    \label{eq:inverse_laplace}
     \sum_{m = 1}^{n} (-1)^{m - 1} 2 \mathbf{Im}(\Omega_{r^{m}\beta}b_{ll'\dots \beta rr' \dots \overline{r^{m}} \dots}(E)b_{ll'\dots rr'\dots}^{*}(E')))]
\end{equation}

Now, one can use the same initial equations \eqref{eq: simplified_shred_many_body} and insert them into the obtained expression \eqref{eq:inverse_laplace}, in the conjugated parts on the right. After summing over all indices (energy levels) $(ll' \dots rr' \dots)$ \textbf{(\textit{without accounting for ordering by utilizing the Fermi nature of the wave function, and as a result, getting repeated contributions from the upper definition of density operator terms, which will result in a combinatorial factor})}, we find that \textbf{\textit{all terms on the right that are "non-diagonal", in other words, those that don't have the same sets of indices related to the collector and emitter degrees of freedom in non-conjugated and conjugated terms, after summation, will end up being zero, which arises from the same "2-pole" logic}}, explained earlier in the derivation of the simplification of the "spectral" equation parts. \textbf{\textit{For terms with the same sets of indices, we can utilize the same Sokhotski - Plemelj theorem to reduce the integration over the extra $r^{m}$ energy variable}}. Finally, noting that $(E - E')$ after the inverse transform will result in a time derivative, we get:

\[
    (i\frac{d\sigma_{00}^{n}}{dt} + i(\Gamma_{L\alpha} + \Gamma_{L\beta})\sigma_{00}^{n})(n!)^{2} = i\Gamma_{R\alpha}\sigma_{\alpha \alpha}^{n - 1} (n)! * (n - 1)! * n
\]

\[
     + i\Gamma_{R\beta}\sigma_{\beta \beta}^{n - 1} (n)! * (n - 1)! * n + i\Gamma_{R\alpha\beta}\sigma_{\alpha\beta}^{n - 1} (n)! * (n - 1)! * n 
\]

\begin{equation}
   + i\Gamma_{R\alpha\beta}\sigma_{\beta\alpha}^{n - 1} (n)! * (n - 1)! * n
\end{equation}

As we see, combinatorial factor can be reduced. Finally, we get:

\begin{equation}
\begin{gathered}
    \frac{d\sigma_{00}^{(n)}}{dt} = \Gamma_{R\alpha}\sigma_{\alpha \alpha}^{(n - 1)} + \Gamma_{R\beta}\sigma_{\beta\beta}^{(n - 1)} - (\Gamma_{L\alpha} + \Gamma_{L\beta})\sigma_{00}^{(n)} \\
    + \Gamma_{R\alpha\beta}\sigma_{\alpha\beta}^{(n-1)} + \Gamma_{R\alpha\beta}\sigma_{\beta\alpha}^{(n-1)}
\end{gathered}
\end{equation}

Applying analogous considerations, we can derive all of the remaining equations (note that all density operator elements corresponding to different numbers of electrons in the "2-D bottleneck" will be zero, for example, $\sigma_{\alpha(\alpha\beta)}$, because of total charge conservation in our P-C system):

\begin{equation}
\begin{gathered}
    \frac{d\sigma_{(\alpha\beta)(\alpha\beta)}^{(n)}}{dt} = \Gamma_{L\alpha}\sigma_{\beta\beta}^{(n)} + \Gamma_{L\beta}\sigma_{\alpha\alpha}^{(n)} - (\Gamma_{R\alpha} + \Gamma_{R\beta})\sigma_{(\alpha\beta)(\alpha\beta)}^{(n)} \\
    - \Gamma_{L\alpha\beta}\sigma_{\alpha\beta}^{(n)} -\Gamma_{L\alpha\beta}\sigma_{\beta\alpha}^{(n)}
\end{gathered}
\end{equation}

\begin{equation}
\begin{gathered}
    \frac{d\sigma_{\alpha\alpha}^{(n)}}{dt} = \Gamma_{L\alpha}\sigma_{00}^{(n)} + \Gamma_{R\beta}\sigma_{(\alpha\beta)(\alpha\beta)}^{(n - 1)} - \Gamma_{R\alpha}\sigma_{\alpha\alpha}^{(n)} - \Gamma_{L\beta}\sigma_{\alpha\alpha}^{(n)} \\
     + \frac{\Delta\Gamma}{2}(\sigma_{\alpha\beta}^{(n)} + \sigma_{\beta\alpha}^{(n)}) 
\end{gathered}
\end{equation}

\begin{equation}
\begin{gathered}
    \frac{d\sigma_{\beta\beta}^{(n)}}{dt} = \Gamma_{L\beta}\sigma_{00}^{(n)} + \Gamma_{R\alpha}\sigma_{(\alpha\beta)(\alpha\beta)}^{(n - 1)} - \Gamma_{L\alpha}\sigma_{\beta\beta}^{(n)} - \Gamma_{R\beta}\sigma_{\beta\beta}^{(n)} \\
     + \frac{\Delta\Gamma}{2}(\sigma_{\alpha\beta}^{(n)} + \sigma_{\beta\alpha}^{(n)}) 
\end{gathered}
\end{equation}

\begin{equation}
\begin{gathered}
    \frac{d\sigma_{\alpha\beta}^{(n)}}{dt} = \Gamma_{L\alpha\beta}\sigma_{00}^{(n)} + \Gamma_{R\alpha\beta}\sigma_{(\alpha\beta)(\alpha\beta)}^{(n - 1)} + \frac{\Delta\Gamma}{2}(\sigma_{\alpha\alpha}^{(n)} + \sigma_{\beta\beta}^{(n)}) \\
    - \frac{(\Gamma_{R\alpha} + \Gamma_{R\beta} + \Gamma_{L\alpha} + \Gamma_{L\beta})}{2}\sigma_{\alpha\beta}^{(n)} + i(E_{\beta} - E_{\alpha})\sigma_{\alpha\beta}^{(n)}
\end{gathered}
\end{equation}

Further we use notation $\Gamma = \frac{\Gamma_{R\alpha} + \Gamma_{R\beta} + \Gamma_{L\alpha} + \Gamma_{L\beta}}{2} + i(E_{\alpha} - E_{\beta})$.

\begin{widetext}

Now, we can present obtained ode's in a Master-equations like manner:

\begin{equation}
    \dot\sigma^{n} = A\sigma^{n} + B\sigma^{n - 1}
\end{equation}

where:

\begin{equation}
\begin{gathered}
A = \begin{pmatrix}
-(\Gamma_{L\alpha} + \Gamma_{L\beta}) & 0 & 0 & 0 & 0 & 0 \\
\Gamma_{L\alpha} & -(\Gamma_{R\alpha} + \Gamma_{L\beta}) & 0 & \frac{\Delta\Gamma}{2} & \frac{\Delta\Gamma}{2} & 0 \\
\Gamma_{L\beta}& 0 & -(\Gamma_{R\beta} + \Gamma_{L\alpha}) & \frac{\Delta\Gamma}{2} & \frac{\Delta\Gamma}{2} & 0 \\
\Gamma_{L\alpha\beta} & \frac{\Delta\Gamma}{2} & \frac{\Delta\Gamma}{2} & -\Gamma & 0 & 0 \\ 
\Gamma_{L\alpha\beta} & \frac{\Delta\Gamma}{2} & \frac{\Delta\Gamma}{2} & 0 & -\Gamma^{*} & 0 \\ 
0 & \Gamma_{L\beta} & \Gamma_{L\alpha} & -\Gamma_{L\alpha\beta} & -\Gamma_{L\alpha\beta} & -(\Gamma_{R\alpha} + \Gamma_{R\beta})
\end{pmatrix} \\
\end{gathered}
\end{equation}

\begin{equation}
\begin{gathered}
B = \begin{pmatrix}
0 & \Gamma_{R\alpha} & \Gamma_{R\beta} & \Gamma_{R\alpha\beta} & \Gamma_{R\alpha\beta} & 0 \\
0 & 0 & 0 & 0 & 0 & \Gamma_{R\beta} \\
0 & 0 & 0 & 0 & 0 & \Gamma_{R\alpha} \\
0 & 0 & 0 & 0 & 0 & \Gamma_{R\alpha\beta} \\
0 & 0 & 0 & 0 & 0 & \Gamma_{R\alpha\beta} \\
0 & 0 & 0 & 0 & 0 & 0 \\
\end{pmatrix}
\end{gathered}
\end{equation}

\end{widetext}

Now, we can use MacDonald's formula \cite{McDonald} to derive current's noise power spectrum to study it's statistics and special attributes:

\begin{widetext}

\begin{equation}
    S(\omega) = 2e^{2}\omega\int_{0}^{+\infty}\sum_{n = 0}^{+\infty}n^{2}\frac{d(\sigma^{n}_{00} + \sigma^{n}_{\alpha\alpha} + \sigma^{n}_{\beta\beta} + \sigma^{n}_{(\alpha\beta)(\alpha\beta)})}{dt}\sin(\omega t)dt
\end{equation}

\end{widetext}

We can effectively do the trace above by transition to vector form, complex fourier transform, and use of previous equations:

\begin{equation}
\begin{gathered}
    \int_{0}^{+\infty}\sum_{n = 0}^{+\infty}n^{2}\frac{d\sigma^{(n)}}{dt}e^{i\omega t}dt = (A + B)\int_{0}^{+\infty}\sum_{n = 0}^{+\infty}n^{2}\sigma^{(n)} e^{i\omega t}dt \\
    + 2B\int_{0}^{+\infty}\sum_{n = 0}^{+\infty}n\sigma^{(n)} e^{i\omega t}dt + B\int_{0}^{+\infty}\sum_{n = 0}^{+\infty}\sigma^{(n)} e^{i\omega t}dt
\end{gathered}
\end{equation}

Conservation of trace for full density operator results in zero income from the first term, $(A + B)_{00} + (A + B)_{\alpha\alpha} + (A + B)_{\beta\beta} + (A + B)_{(\alpha\beta)(\alpha\beta)} = 0$.

After use of some fouirer transform properties, we get:

\begin{equation}
\begin{gathered}
    2B\int_{0}^{+\infty}\sum_{n = 0}^{+\infty}n\sigma^{(n)} e^{i\omega t}dt + B\int_{0}^{+\infty}\sum_{n = 0}^{+\infty}\sigma^{(n)} e^{i\omega t}dt = \\
    \hspace{1cm}(2(B(A + B + i\omega)^{-1})^{2} - B(A + B + i\omega)^{-1})\sigma(\infty)
\end{gathered}
\end{equation}
 
where $\sigma(\infty)$ is the stationary limit:

\begin{equation}
    (A + B)\sigma(\infty) = 0
\end{equation}

Now, we can study different limit cases for $S(\omega)$.

\subsection{In - coherent case}

We call In - coherent such P-C structures, which will result in $\Gamma_{R\alpha\beta} = \Gamma_{L\alpha\beta} = 0$. We also study only the symmetric case, for any system further. So, we get:

\begin{equation}
    \sigma(\infty) = \begin{pmatrix}
\frac{1}{4} \\
\frac{1}{4} \\ 
\frac{1}{4} \\
0\\
0\\
\frac{1}{4}

\end{pmatrix}
\end{equation}

In the fig. 6 one can see illustration of $S(\omega)$ in this limit for different values of $\Gamma$.

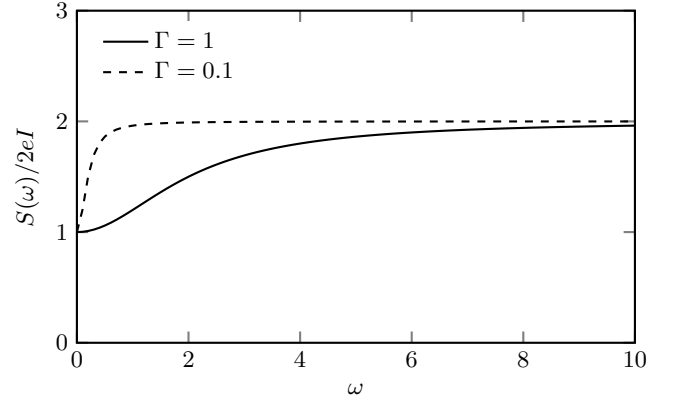
\begin{figure}[H]
\begin{tikzpicture}
    \begin{axis}[
        width=9cm, height=6cm, % Size of the plot
        xlabel={$\omega$}, % X-axis label
        ylabel={$S(\omega)/2eI$}, % Y-axis label
        xmin=0, xmax=10, % X-axis range
        ymin=0.0, ymax=3, % Y-axis range
        axis line style={thick}, % Thick axis lines
        tick style={thick}, % Thick ticks
        legend style={draw=none, cells={anchor=west}},
        legend pos=north west % Position of legend
    ]
    
    % Plot the solid line
    \addplot [
        thick,
        domain=0:10,
        samples=100,
        color=black,
    ] {(4 + 2*x^2) / (4 + x^2)};
    \addlegendentry{$\Gamma = 1$}

    % Plot the dashed line
    \addplot [
        dashed,
        thick,
        domain=0:10,
        samples=100,
        color=black,
    ] {(0.04 + 2*x^2) / (0.04 + x^2)};
    \addlegendentry{$\Gamma = 0.1$}

    \end{axis}
\end{tikzpicture}
\caption{Noise power spectrum in in-coherent case.}
\end{figure}

\begin{equation}
    \frac{S(\omega)}{2eI} = \frac{4\Gamma^2 + 2\omega^2}{4\Gamma^2 + \omega^2}
\end{equation}

The key point in this limit is that the noise does not depend on $\Delta = E_{\alpha} - E_{\beta}$, or any other special characteristics of the "2D bottleneck" in the P-C contact, which fully coincides with the "non-coherent" regime of current. In the high-frequency limit, we obtain classical "Schottky" noise.

\subsection{Coherent case}

The Exact result of the noise for arbitrary $\Gamma_{R, L(\alpha\beta)}$ can be obtained; however, its full analytical expression is quite lengthy.  Nevertheless, we are still able to capture the key aspects in some limiting cases. 
In the limit of "low coherence," $\Delta \ll \Gamma$, we get:

\begin{equation}
    \sigma(\infty) = \begin{pmatrix}
\frac{1}{3} \\
\frac{1}{6} \\ 
\frac{1}{6} \\
\frac{1}{6}\\
\frac{1}{6}\\
0

\end{pmatrix}
\end{equation}

\begin{equation}
    \frac{S(\omega)}{2eI} = \frac{8\Gamma^2 + \omega^2}{16\Gamma^2 + \omega^2}
\end{equation}

As we can see, it looks the same as in the "in-coherent" case. In the limit of $\Delta \gg \Gamma$, and neglecting low-order corrections in the stationary density operator solution:

\begin{equation}
    \sigma(\infty) = \begin{pmatrix}
\frac{1}{4} \\
\frac{1}{4} \\ 
\frac{1}{4} \\
0\\
0\\
\frac{1}{4}

\end{pmatrix}
\end{equation}

\begin{equation}
\begin{gathered}
    \frac{S(\omega)}{2eI} = \frac{ 2(64\Gamma^6 + \omega^2(\omega^2 + 8\Gamma^2)^2 + \Delta^4(2 \Gamma^2 + \omega^2))}{3(4 \Gamma^2 + \omega^2) ( 
   4 \Gamma^2 + (\omega - \Delta)^2) ( 4 \Gamma^2 + (\omega + \Delta)^2)}\\
    + \frac{2\Delta^2 (24 \Gamma^4 + 6 \Gamma^2 \omega^2 - 2 \omega^4)}{3(4 \Gamma^2 + \omega^2)( 
   4 \Gamma^2 + (\omega - \Delta)^2) ( 4 \Gamma^2 + (\omega + \Delta)^2)}
\end{gathered}
\end{equation}

\begin{figure}
\begin{tikzpicture}
    \begin{axis}[
        width=9cm, height=6cm, % Size of the plot
        xlabel={$\omega$}, % X-axis label
        ylabel={$S(\omega)/2eI$}, % Y-axis label
        xmin=0, xmax=100, % X-axis range
        ymin=0.0, ymax=2, % Y-axis range
        axis line style={thick}, % Thick axis lines
        tick style={thick}, % Thick ticks
        legend style={draw=none, cells={anchor=west}},
        legend pos=north west % Position of legend
    ]
    
    % Plot the solid line
    \addplot [
        thick,
        domain=0:100,
        samples=200,
        color=black,
    ] {2 * (64 + x^2 * (x^2 + 8)^2 + (25)^4 * (2 + x^2) + (25)^2 * (24 + 6 * x^2 - 2 * x^4)) / (3 * (4 + x^2) * (4 + (x - 25)^2) * (4 + (x + 25)^2))};
    \addlegendentry{$\Gamma = 1, \Delta = 25$}

    % Plot the dashed line
    \addplot [
        dashed,
        domain=0:100,
        samples=100,
        color=black,
    ] {(x^2 + 8) / (x^2 + 16)};
    \addlegendentry{$\Gamma = 1, \Delta = 0$}

    \end{axis}

    \end{tikzpicture}
\caption{Noise power spectrum in coherent case.}
\end{figure}
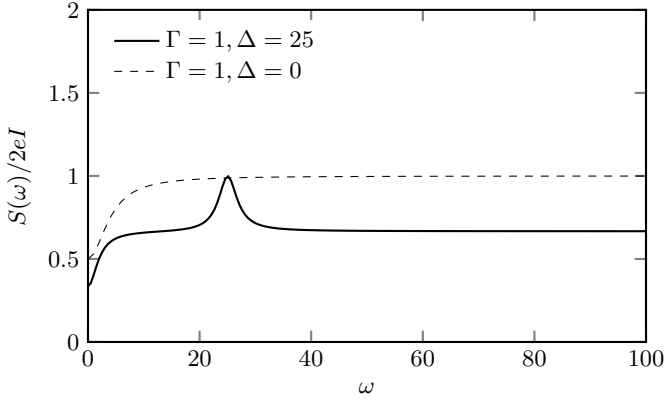

In the fig. 7 one can see illustration of $S(\omega)$ in this limit for different values of $\Delta$.

In the $\Gamma \ll \Delta$ limit, there is a sharp peak at the $\Delta$ frequency, which corresponds to the "coherent limit" of the P-C contact. Using this representation of the noise power spectrum, one can, for example, determine $\Delta$ by analyzing the noise of the total current.

\section{\label{sec:level1} PC 2-d "bottleneck" near quantum dot qubit}

Now, in order to describe our P-C as a quantum measurement device, we introduce a qubit into our system. The transition rates in the P-C now become dependent on the state of the qubit due to Coulomb interaction. We can work within the same formalism as before, accounting for the fact that our Schrödinger multiparticle state now has an additional index: $x$ for the $|0\rangle$ state of the qubit and $y$ for the $|1\rangle$ state.

For the $|0\rangle$ state, the transition amplitudes are $\Omega_{l, r, (\alpha, \beta)}$, while for the $|1\rangle$ state, they undergo a constant change (in our approximation): $\Omega_{l, r, (\alpha, \beta)}^{'} = \Omega_{l, r, (\alpha, \beta)} + \delta\Omega$. Applying an analogous technique, we obtain the following equations for the reduced density operator of the qubit-bottleneck system:

\begin{equation}
    \dot\sigma^{n} = A_{q}\sigma^{n} + B_{q}\sigma^{n - 1}
\end{equation}

where:

\begin{equation}
\begin{gathered}
    A_{q} = \begin{pmatrix}
A & i\Omega_{0} & -i\Omega_{0} & 0 \\
i\Omega_{0} & A^{'(xy)} & 0 & -i\Omega_{0}\\
-i\Omega_{0} & 0 & A^{'(yx)} & i\Omega_{0} \\
0 & -i\Omega_{0} & i\Omega_{0} & A^{\delta} \\
\end{pmatrix} \\
B_{q} = \begin{pmatrix}
B & 0 & 0 & 0 \\
0 & B^{'(xy)} & 0 & 0\\
0 & 0 & B^{'(yx)} & 0 \\
0 & 0 & 0 & B^{\delta} \\
\end{pmatrix}
\end{gathered}
\end{equation}

\begin{widetext}

\vspace{1cm}
\begin{equation}
    A^{'(xy)} = A + \begin{pmatrix}
-\delta\frac{(\Gamma_{L\alpha} + \Gamma_{L\beta})}{2} & 0 & 0 & 0 & 0 & 0\\
\delta\Gamma_{L\alpha} & -\delta\frac{(\Gamma_{R\alpha} + \Gamma_{L\beta})}{2} & 0 & \frac{\Gamma_{\Delta}^{'} - \Gamma_{\Delta}}{2} & 0 & 0\\
\delta\Gamma_{L\beta} & 0 & -\delta\frac{(\Gamma_{R\beta} + \Gamma_{L\alpha})}{2} & 0 & \frac{\Gamma_{\Delta}^{'} - \Gamma_{\Delta}}{2} & 0 \\
\delta\Gamma_{L\alpha}  & \frac{\Gamma_{\Delta}^{'} - \Gamma_{\Delta}}{2} & 0 & -\delta\frac{(\Gamma_{R\alpha} + \Gamma_{L\beta})}{2} & 0 & 0 \\
\delta\Gamma_{L\beta}  & 0 & \frac{\Gamma_{\Delta}^{'} - \Gamma_{\Delta}}{2} & 0 & -\delta\frac{(\Gamma_{R\beta} + \Gamma_{L\alpha})}{2} & 0 \\
0 & \delta\Gamma_{L\beta} & \delta\Gamma_{L\alpha} & -\delta\Gamma_{L\beta} & -\delta\Gamma_{L\alpha} & -\delta\frac{(\Gamma_{R\alpha} + \Gamma_{R\beta})}{2} \\
\end{pmatrix} + i\epsilon
\end{equation}

\newpage

\begin{equation}
    A^{'(yx)} = A + \begin{pmatrix}
-\delta\frac{(\Gamma_{L\alpha} + \Gamma_{L\beta})}{2} & 0 & 0 & 0 & 0 & 0\\
\delta\Gamma_{L\alpha} & -\delta\frac{(\Gamma_{R\alpha} + \Gamma_{L\beta})}{2} & 0 & 0 & \frac{\Gamma_{\Delta}^{'} - \Gamma_{\Delta}}{2} & 0\\
\delta\Gamma_{L\beta} & 0 & -\delta\frac{(\Gamma_{R\beta} + \Gamma_{L\alpha})}{2} & \frac{\Gamma_{\Delta}^{'} - \Gamma_{\Delta}}{2} & 0 & 0 \\
\delta\Gamma_{L\beta}  & 0 & \frac{\Gamma_{\Delta}^{'} - \Gamma_{\Delta}}{2} & -\delta\frac{(\Gamma_{R\beta} + \Gamma_{L\alpha})}{2} & 0 & 0 \\
\delta\Gamma_{L\alpha}  & \frac{\Gamma_{\Delta}^{'} - \Gamma_{\Delta}}{2} & 0 & 0 & -\delta\frac{(\Gamma_{R\alpha} + \Gamma_{L\beta})}{2} & 0 \\
0 & \delta\Gamma_{L\beta} & \delta\Gamma_{L\alpha} & -\delta\Gamma_{L\alpha} & -\delta\Gamma_{L\beta} & -\delta\frac{(\Gamma_{R\alpha} + \Gamma_{R\beta})}{2} \\
\end{pmatrix} - i\epsilon
\end{equation}

\normalsize

\begin{equation}
\begin{gathered}
    B^{'(xy)} = B + \begin{pmatrix}
0 & \delta\Gamma_{R\alpha} & \delta\Gamma_{R\beta}  & \delta\Gamma_{R\alpha}  & \delta\Gamma_{R\beta}  & 0\\
0 & 0 & 0 & 0 & 0 & \delta\Gamma_{R\beta}\\
0 & 0 & 0 & 0 & 0 & \delta\Gamma_{R\alpha}\\
0 & 0 & 0 & 0 & 0 & \delta\Gamma_{R\beta}\\
0 & 0 & 0 & 0 & 0 & \delta\Gamma_{R\alpha}\\
0 & 0 & 0 & 0 & 0 & 0\\
\end{pmatrix} \\
    B^{'(yx)} = B + \begin{pmatrix}
0 & \delta\Gamma_{R\alpha} & \delta\Gamma_{R\beta}  & \delta\Gamma_{R\beta}  & \delta\Gamma_{R\alpha}  & 0\\
0 & 0 & 0 & 0 & 0 & \delta\Gamma_{R\beta}\\
0 & 0 & 0 & 0 & 0 & \delta\Gamma_{R\alpha}\\
0 & 0 & 0 & 0 & 0 & \delta\Gamma_{R\alpha}\\
0 & 0 & 0 & 0 & 0 & \delta\Gamma_{R\beta}\\
0 & 0 & 0 & 0 & 0 & 0\\
\end{pmatrix}
\end{gathered}
\end{equation}

\end{widetext}

Here, we use $'$ symbol for width constants of energy levels which are calculated with a $\delta\Omega$ shift, $\delta\Gamma_{L, R, (\alpha, \beta)} = \pi\rho_{L, R}\delta\Omega\cdot\Omega_{L, R, (\alpha, \beta)}$, $\delta(\Gamma_{\alpha} + \Gamma_{\beta}) = (\Gamma^{'}_{\alpha} + \Gamma^{'}_{\beta} - \Gamma_{\alpha} - \Gamma_{\beta})$, and $\epsilon = E_{1} - E_{2}$ is the difference between energy levels of two dots, representing states in our qubit. 

\subsection{Current noise power-spectrum in presence of qubit}

All previous derivations of the formula for the noise power spectrum remain valid with the transition $A, B \rightarrow A_{q}, B_{q}$. However, due to the presence of the qubit in the system, the density operator now contains 24 elements. As a result, the analytical treatment of $S(\omega)$ and even $\sigma(\infty)$ becomes highly challenging, even with the use of specialized symbolic computational software.

Therefore, we develop methods for the effective analysis of $S(\omega)$ and $\sigma(\infty)$ in different limits, with respect to $\Gamma$ and $\Omega_{0}$.

\subsubsection{Stationary state approximation}

For calculating $S(\omega)$, we first need to obtain appropriate form of $\sigma(\infty)$. It is defined as before:

\begin{equation}
    (A_{q} + B_{q})\sigma(\infty) = 0
\end{equation}

Looking at the form of $A_{q}, B_{q}$ we can rewrite this problem in terms of block operators:

\begin{widetext}

\begin{equation}
    \begin{pmatrix}
C_{xx} & i\Omega_{0} & -i\Omega_{0} & 0 \\
i\Omega_{0} &  C_{xy}& 0 & -i\Omega_{0}\\
-i\Omega_{0} & 0 & C_{yx} & i\Omega_{0} \\
0 & -i\Omega_{0} & i\Omega_{0} & C_{yy} \\
\end{pmatrix} 
\begin{pmatrix}
\rho_{xx} \\
\rho_{xy}\\
\rho_{yx}\\
\rho_{yy}\\
\end{pmatrix} = 0
\end{equation}

\end{widetext}

where $\rho_{x, y}$ are 6-dimensional vectors representing different parts of the density operator with respect to the qubit degrees of freedom, and $C_{x, y} = \mathcal{O}(\Gamma)$.

In general, we can consider two key limits: $\Omega_{0} \ll \Gamma$ and $\Omega_{0} \gg \Gamma$.

For the first case, the calculation is straightforward: we can expand the solution in a series of $\Omega_{0}$, as it is often done in traditional perturbation theory, and subsequently compute each term. In the zero-order approximation:

\begin{equation}
    \begin{pmatrix}
C_{xx} & 0 & 0 & 0 \\
0 &  C_{xy}& 0 & 0\\
0 & 0 & C_{yx} & 0 \\
0 & 0 & 0 & C_{yy} \\
\end{pmatrix} \begin{pmatrix}
\rho_{xx} \\
\rho_{xy}\\
\rho_{yx}\\
\rho_{yy}\\
\end{pmatrix}^{0} = 0
\end{equation}

Here, $C_{xx}$ and $C_{yy}$ matrices are $A + B$ and $A^{'} + B^{'}$, while $C_{xy}$ and $C_{yx}$ correspond to $A^{'(xy)} + B^{'(xy)}$ and $A^{'(yx)} + B^{'(yx)}$. The corresponding null-space 24-dimensional problem simplifies to four 6-dimensional problems.

Additionally, we note that $C_{xy}$ and $C_{yx}$ do not have any solutions for the equation $C_{xy} \rho_{xy} = 0$, since their determinant is always nonzero in non-degenerate cases. This result is fully reasonable, as in the zero-order approximation, where there is no tunneling in the qubit, there will be no antidiagonal elements in the reduced density operator for the qubit.

\vspace{10pt}

To obtain the first-order approximation, we expand the equation in a series and derive the equation:

\begin{equation}
    \begin{gathered}
    \begin{pmatrix}
C_{xx} & 0 & 0 & 0 \\
0 &  C_{xy}& 0 & 0\\
0 & 0 & C_{yx} & 0 \\
0 & 0 & 0 & C_{yy} \\
\end{pmatrix} \begin{pmatrix}
\rho_{xx} \\
\rho_{xy}\\
\rho_{yx}\\
\rho_{yy}\\
\end{pmatrix}^{\Omega_{0}} \\ + \begin{pmatrix}
0 & i\Omega_{0} & -i\Omega_{0} & 0 \\
i\Omega_{0} &  0 & 0 & -i\Omega_{0}\\
-i\Omega_{0} & 0 & 0 & i\Omega_{0} \\
0 & -i\Omega_{0} & i\Omega_{0} & 0 \\
\end{pmatrix} 
\begin{pmatrix}
\rho_{xx} \\
\rho_{xy}\\
\rho_{yx}\\
\rho_{yy}\\
\end{pmatrix}^{0} = 0
 \end{gathered}
\end{equation}

\normalsize

which in our case simplifies to equations:
\begin{equation}
\begin{gathered}
    C_{xx}\rho_{xx}^{\Omega_{0}} = 0\\
    C_{xy}\rho_{xy}^{\Omega_{0}} = i\Omega_{0}(\rho_{yy}^{0} - \rho_{xx}^{0})\\
    C_{yx}\rho_{yx}^{\Omega_{0}} = i\Omega_{0}(\rho_{xx}^{0} - \rho_{yy}^{0})\\
    C_{yy}\rho_{xx}^{\Omega_{0}} = 0
\end{gathered}
\end{equation}

Non-zero solutions of first and last equation dont give new linear independent stationary states, so, finally, we get:
\begin{equation}
    \begin{pmatrix}
\rho_{xx} \\
\rho_{xy}\\
\rho_{yx}\\
\rho_{yy}\\
\end{pmatrix}^{\Omega_{0}} =  i\Omega_{0}\begin{pmatrix}
0 \\
C_{xy}^{-1}(\rho_{yy}^{0} - \rho_{xx}^{0})\\
C_{yx}^{-1}(\rho_{xx}^{0} - \rho_{yy}^{0})\\
0\\
\end{pmatrix}
\end{equation}

The inverse operators involved in this calculation can have an adequate analytical form, at least under certain assumptions about the conditions of our problem, which we will discuss further. This calculation can be extended to obtain corrections of order $\Omega_{0}^2$ and higher. In fact, nonzero corrections for the diagonal elements of the qubit density operator appear only from the second order onward. With each even order, we obtain corrections to the diagonal elements, while with each odd order, corrections appear for the antidiagonal elements. This makes sense since we are expanding in terms of $i\Omega_{0}$.

\vspace{10pt}

Now, consider the opposite limit: $\Omega_{0} \gg \Gamma$. To perform a perturbative expansion, we cannot directly exploit the initial form of our equation. Instead, we can reformulate the problem by noting that we can express $\rho_{yx}$ and $\rho_{yy}$ in terms of $\rho_{xx}$ and $\rho_{xy}$ by using the first and second rows of our block equations:
\begin{equation}
\begin{gathered}
    \rho_{yx} = \rho_{xy} - \frac{i}{\Omega_{0}}C_{xx}\rho_{xx}\\
    \rho_{yy} = \rho_{xx} - \frac{i}{\Omega_{0}}C_{xy}\rho_{xy}
\end{gathered}
\end{equation}

inserting those equations in last two rows, we can finally write:
\begin{equation}
    \begin{pmatrix}
-\frac{i}{\Omega_{0}}C_{yx}C_{xx} & C_{xy} + C_{yx} \\
C_{xx} + C_{yy} & -\frac{i}{\Omega_{0}}C_{yy}C_{xy}\\
\end{pmatrix} \begin{pmatrix}
\rho_{xx} \\
\rho_{xy}\\
\end{pmatrix} = 0
\end{equation}

This form gives us ability to again expand solutions in terms of new perturbation parameter $\frac{i}{\Omega_{0}}$. For the zero order approximation:
\begin{equation}
    \begin{pmatrix}
0 & C_{xy} + C_{yx} \\
C_{xx} + C_{yy} & 0\\
\end{pmatrix} \begin{pmatrix}
\rho_{xx} \\
\rho_{xy}\\
\end{pmatrix}^{0} = 0
\end{equation}

again, one can see, that $C_{xy} + C_{yx}$ has an empty nullspace. Thus, solution is given by kernel of $C_{xx} + C_{yy}$:
\begin{equation}
    \begin{pmatrix}
\rho_{xx} \\
\rho_{xy}\\
\rho_{yx}\\
\rho_{yy}\\
\end{pmatrix}^{0} = \begin{pmatrix}
\ker(C_{xx} + C_{yy})\\
0\\
0\\
\ker(C_{xx} + C_{yy})\\
\end{pmatrix}  
\end{equation}

Analogous to the first limit, we can calculate the first order correction:
\begin{equation}
    \begin{gathered}
    \begin{pmatrix}
0 & C_{xy} + C_{yx} \\
C_{xx} + C_{yy} & 0\\
\end{pmatrix} \begin{pmatrix}
\rho_{xx} \\
\rho_{xy}\\
\end{pmatrix}^{\frac{i}{\Omega_{0}}} \\ +\begin{pmatrix}
-\frac{i}{\Omega_{0}}C_{yx}C_{xx} & 0 \\
0 & -\frac{i}{\Omega_{0}}C_{yy}C_{xy}\\
\end{pmatrix} \begin{pmatrix}
\rho_{xx} \\
\rho_{xy}\\
\end{pmatrix}^{0} = 0
\end{gathered}
\end{equation}

\normalsize

this results in:
\begin{equation}
    \begin{pmatrix}
\rho_{xx} \\
\rho_{xy}\\
\end{pmatrix}^{\frac{i}{\Omega_{0}}} = \frac{i}{\Omega_{0}}\begin{pmatrix}
0 \\
(C_{xy} + C_{yx})^{-1}C_{yx}C_{xx}\rho_{xx}^{0} \\
\end{pmatrix}
\end{equation}

and finally, full first order correction:
\begin{equation}
    \begin{pmatrix}
\rho_{xx} \\
\rho_{xy} \\
\rho_{yx} \\
\rho_{yy} \\
\end{pmatrix}^{\frac{i}{\Omega_{0}}} = \frac{i}{\Omega_{0}}\begin{pmatrix}
0 \\
(C_{xy} + C_{yx})^{-1}C_{yx}C_{xx}\rho_{xx}^{0} \\
-(C_{xy} + C_{yx})^{-1}C_{xy}C_{xx}\rho_{xx}^{0} \\
\rho_{yy} \\
0 \\
\end{pmatrix}
\end{equation}

We again see same pattern for even and odd orders of correction.

From now on, we will consider only the first order correction in all calculations.

For the $\Omega_{0} \ll \Gamma$ limit we get:
\begin{equation}
    \begin{pmatrix}
\rho_{xx} \\
\rho_{xy} \\
\rho_{yx} \\
\rho_{yy} \\
\end{pmatrix}^{0} = \begin{pmatrix}
\rho_{xx}^{0} \\
0 \\
0 \\
\rho_{yy}^{0} \\
\end{pmatrix}
\end{equation}

\begin{equation}
    \rho_{xx}^{0} = \begin{pmatrix}
1 \\
1 \\
1 \\
\frac{2 \Gamma_{ab}}{\Gamma_{a} + \Gamma_{b}} \\
\frac{2 \Gamma_{ab}}{\Gamma_{a} + \Gamma_{b}} \\
1 \\
\end{pmatrix}
\end{equation}

\begin{equation}
    \rho_{yy}^{0} = \begin{pmatrix}
1 \\
1 \\
1 \\
\frac{2 \Gamma_{ab}^{'}}{\Gamma_{a}^{'} + \Gamma_{b}^{'}} \\
\frac{2 \Gamma_{ab}^{'}}{\Gamma_{a}^{'} + \Gamma_{b}^{'}} \\
1 \\
\end{pmatrix}
\end{equation}

\begin{equation}
    \begin{pmatrix}
\rho_{xx} \\
\rho_{xy} \\
\rho_{yx} \\
\rho_{yy} \\
\end{pmatrix}^{\Omega_{0}} = i\Omega_{0}\begin{pmatrix}
0 \\
\rho_{xy}^{\Omega_{0}} \\
\rho_{yx}^{\Omega_{0}} \\
0 \\
\end{pmatrix}
\end{equation}

\begin{equation}
    \rho_{xy}^{\Omega_{0}} = -\begin{pmatrix}
\frac{\Gamma_{\alpha\beta}(3\Gamma_{\alpha}\delta\Gamma_{\alpha} + \Gamma_{\alpha}\delta\Gamma_{\beta} + \Gamma_{\beta}\delta\Gamma_{\alpha} + 3\Gamma_{\beta}\delta\Gamma_{\beta})}{2\Gamma_{\alpha}\Gamma_{\beta}(\Gamma_{\alpha} + \Gamma_{\beta})^2} \\
\frac{\Gamma_{\alpha\beta}(3\Gamma_{\alpha}\delta\Gamma_{\alpha} + \Gamma_{\alpha}\delta\Gamma_{\beta} - 3\Gamma_{\beta}\delta\Gamma_{\alpha} - \Gamma_{\beta}\delta\Gamma_{\beta})}{2\Gamma_{\alpha}\Gamma_{\beta}(\Gamma_{\alpha} + \Gamma_{\beta})^2}\\
\frac{\Gamma_{\alpha\beta}(-\Gamma_{\alpha}\delta\Gamma_{\alpha} -3 \Gamma_{\alpha}\delta\Gamma_{\beta} + \Gamma_{\beta}\delta\Gamma_{\alpha} + 3\Gamma_{\beta}\delta\Gamma_{\beta})}{2\Gamma_{\alpha}\Gamma_{\beta}(\Gamma_{\alpha} + \Gamma_{\beta})^2} \\
-\Gamma_{\alpha\beta}\frac{4 (\delta\Gamma_{\alpha} + \delta\Gamma_{\beta})}{
   (\Gamma_{\alpha} + \Gamma_{\beta})^3}
+ \frac{2(\delta\Gamma_{\alpha} + \delta\Gamma_{\beta})}{(\Gamma_{\alpha} + \Gamma_{\beta})^2} \\
-\Gamma_{\alpha\beta}\frac{4 (\delta\Gamma_{\alpha} + \delta\Gamma_{\beta})}{
   (\Gamma_{\alpha} + \Gamma_{\beta})^3}
+ \frac{2(\delta\Gamma_{\alpha} + \delta\Gamma_{\beta})}{(\Gamma_{\alpha} + \Gamma_{\beta})^2} \\
-\frac{\Gamma_{\alpha\beta}(\Gamma_{\alpha}\delta\Gamma_{\alpha} + 3\Gamma_{\alpha}\delta\Gamma_{\beta} + 3\Gamma_{\beta}\delta\Gamma_{\alpha} + \Gamma_{\beta}\delta\Gamma_{\beta})}{2\Gamma_{\alpha}\Gamma_{\beta}(\Gamma_{\alpha} + \Gamma_{\beta})^2} \\
\end{pmatrix}
\end{equation}

\begin{equation}
    \rho_{yx}^{\Omega_{0}} = \begin{pmatrix}
\frac{\Gamma_{\alpha\beta}(3\Gamma_{\alpha}\delta\Gamma_{\alpha} + \Gamma_{\alpha}\delta\Gamma_{\beta} + \Gamma_{\beta}\delta\Gamma_{\alpha} + 3\Gamma_{\beta}\delta\Gamma_{\beta})}{2\Gamma_{\alpha}\Gamma_{\beta}(\Gamma_{\alpha} + \Gamma_{\beta})^2} \\
\frac{\Gamma_{\alpha\beta}(3\Gamma_{\alpha}\delta\Gamma_{\alpha} + \Gamma_{\alpha}\delta\Gamma_{\beta} - 3\Gamma_{\beta}\delta\Gamma_{\alpha} - \Gamma_{\beta}\delta\Gamma_{\beta})}{2\Gamma_{\alpha}\Gamma_{\beta}(\Gamma_{\alpha} + \Gamma_{\beta})^2}\\
\frac{\Gamma_{\alpha\beta}(-\Gamma_{\alpha}\delta\Gamma_{\alpha} -3 \Gamma_{\alpha}\delta\Gamma_{\beta} + \Gamma_{\beta}\delta\Gamma_{\alpha} + 3\Gamma_{\beta}\delta\Gamma_{\beta})}{2\Gamma_{\alpha}\Gamma_{\beta}(\Gamma_{\alpha} + \Gamma_{\beta})^2} \\
-\Gamma_{\alpha\beta}\frac{4 (\delta\Gamma_{\alpha} + \delta\Gamma_{\beta})}{
   (\Gamma_{\alpha} + \Gamma_{\beta})^3}
+ \frac{2(\delta\Gamma_{\alpha} + \delta\Gamma_{\beta})}{(\Gamma_{\alpha} + \Gamma_{\beta})^2} \\
-\Gamma_{\alpha\beta}\frac{4 (\delta\Gamma_{\alpha} + \delta\Gamma_{\beta})}{
   (\Gamma_{\alpha} + \Gamma_{\beta})^3}
+ \frac{2(\delta\Gamma_{\alpha} + \delta\Gamma_{\beta})}{(\Gamma_{\alpha} + \Gamma_{\beta})^2} \\
-\frac{\Gamma_{\alpha\beta}(\Gamma_{\alpha}\delta\Gamma_{\alpha} + 3\Gamma_{\alpha}\delta\Gamma_{\beta} + 3\Gamma_{\beta}\delta\Gamma_{\alpha} + \Gamma_{\beta}\delta\Gamma_{\beta})}{2\Gamma_{\alpha}\Gamma_{\beta}(\Gamma_{\alpha} + \Gamma_{\beta})^2} \\
\end{pmatrix}
\end{equation}

And for the $\Omega_{0} \gg \Gamma$:
\begin{equation}
    \begin{pmatrix}
\rho_{xx} \\
\rho_{xy} \\
\rho_{yx} \\
\rho_{yy} \\
\end{pmatrix}^{0} = \begin{pmatrix}
\rho_{xx}^{0} \\
0 \\
0 \\
\rho_{yy}^{0} \\
\end{pmatrix}
\end{equation}

\begin{equation}
    \rho_{xx}^{0} = \rho_{yy}^{0} = \begin{pmatrix}
1 + \frac{2\Gamma_{\alpha\beta}(\delta\Gamma_{\alpha} + \delta\Gamma_{\beta})}{\Gamma_{\alpha}\Gamma_{\beta}} \\
1 + \frac{2\Gamma_{\alpha\beta}(\delta\Gamma_{\alpha} + \delta\Gamma_{\beta})}{\Gamma_{\beta}(\Gamma_{\alpha} + \Gamma_{\beta})}\\
1 + \frac{2\Gamma_{\alpha\beta}(\delta\Gamma_{\alpha} + \delta\Gamma_{\beta})}{\Gamma_{\alpha}(\Gamma_{\alpha} + \Gamma_{\beta})} \\
\frac{\delta\Gamma_{\alpha} + \delta\Gamma_{\beta}}{\Gamma_{\alpha} + \Gamma_{\beta}} + \frac{
 2\Gamma_{\alpha\beta} (\Gamma_{\alpha} + 
    \Gamma_{\beta} -\delta\Gamma_{\alpha} - \delta\Gamma_{\beta})}{(\Gamma_{\alpha} + \Gamma_{\beta})^2} \\
\frac{\delta\Gamma_{\alpha} + \delta\Gamma_{\beta}}{\Gamma_{\alpha} + \Gamma_{\beta}} + \frac{
 2\Gamma_{\alpha\beta} (\Gamma_{\alpha} + 
    \Gamma_{\beta} -\delta\Gamma_{\alpha} - \delta\Gamma_{\beta})}{(\Gamma_{\alpha} + \Gamma_{\beta})^2}  \\
1 \\
\end{pmatrix}
\end{equation}

\begin{equation}
    \begin{pmatrix}
\rho_{xx} \\
\rho_{xy} \\
\rho_{yx} \\
\rho_{yy} \\
\end{pmatrix}^{\frac{i}{\Omega_{0}}} = \frac{i}{\Omega_{0}}\begin{pmatrix}
0 \\
\rho_{xy}^{\frac{i}{\Omega_{0}}} \\
\rho_{yx}^{\frac{i}{\Omega_{0}}} \\
\rho_{yy} \\
0 \\
\end{pmatrix}
\end{equation}

\begin{equation}
    \rho_{xy}^{\frac{i}{\Omega_{0}}} = \begin{pmatrix}
-\frac{\Gamma_{\alpha\beta} (\delta\Gamma_{\alpha} + \delta\Gamma_{\beta})}{\Gamma_{\alpha} + \Gamma_{\beta}} \\
0 \\
0 \\
\frac{\Gamma_{\alpha\beta} (\delta\Gamma_{\alpha} + \delta\Gamma_{\beta})}{\Gamma_{\alpha} + \Gamma_{\beta}} - \frac{1}{2}(\delta\Gamma_{\alpha} + \delta\Gamma_{\beta}) \\
\frac{\Gamma_{\alpha\beta} (\delta\Gamma_{\alpha} + \delta\Gamma_{\beta})}{\Gamma_{\alpha} + \Gamma_{\beta}} - \frac{1}{2}(\delta\Gamma_{\alpha} + \delta\Gamma_{\beta}) \\
\frac{\Gamma_{\alpha\beta} (\delta\Gamma_{\alpha} + \delta\Gamma_{\beta})}{\Gamma_{\alpha} + \Gamma_{\beta}} \\
\end{pmatrix}
\end{equation}

\begin{equation}
    \rho_{yx}^{\frac{i}{\Omega_{0}}} = -\begin{pmatrix}
-\frac{\Gamma_{\alpha\beta} (\delta\Gamma_{\alpha} + \delta\Gamma_{\beta})}{\Gamma_{\alpha} + \Gamma_{\beta}} \\
0 \\
0 \\
\frac{\Gamma_{\alpha\beta} (\delta\Gamma_{\alpha} + \delta\Gamma_{\beta})}{\Gamma_{\alpha} + \Gamma_{\beta}} - \frac{1}{2}(\delta\Gamma_{\alpha} + \delta\Gamma_{\beta}) \\
\frac{\Gamma_{\alpha\beta} (\delta\Gamma_{\alpha} + \delta\Gamma_{\beta})}{\Gamma_{\alpha} + \Gamma_{\beta}} - \frac{1}{2}(\delta\Gamma_{\alpha} + \delta\Gamma_{\beta}) \\
\frac{\Gamma_{\alpha\beta} (\delta\Gamma_{\alpha} + \delta\Gamma_{\beta})}{\Gamma_{\alpha} + \Gamma_{\beta}} \\
\end{pmatrix}
\end{equation}

\subsubsection{Noise power spectrum approximation}

Moving to noise power spectrum analysis, we first note:
\begin{equation}
\begin{gathered}
    (2(B_{q}(A_{q} + B_{q} + i\omega)^{-1})^{2} - B_{q}(A_{q} + B_{q} + i\omega)^{-1})\sigma(\infty) = \\
    \frac{i}{\omega}(-2B_{q}(A_{q} + B_{q} + i\omega)^{-1} + I)B_{q}\sigma(\infty) 
\end{gathered}
\end{equation}

This simplification is possible since $\sigma(\infty)$ is in $\ker(A_{q} + B_{q})$. Thus, we get:
\begin{equation}
    \frac{S(\omega)}{2e^2} = \mathbf{Tr}(\mathbf{Re}(-2B_{q}(A_{q} + B_{q} + i\omega)^{-1} + I)B_{q}\sigma(\infty))
\end{equation}

The last problem is the expression $(A_{q} + B_{q} + i\omega)^{-1}$, which cannot be calculated straightforwardly. We can tackle the analytical problem of inversion of $A_{q} + B_{q} + i\omega$ by exploiting the eigenvalue perspective. Recall the form of $A_{q} + B_{q}$:

\begin{equation}
    A_{q} + B_{q} = \begin{pmatrix}
C_{xx} & 0 & 0 & 0 \\
0 & C_{xy} & 0 & 0 \\
0 & 0 & C_{yx} & 0 \\
0 & 0 & 0 & C_{yy} \\
\end{pmatrix} + i\Omega_{0}\begin{pmatrix}
0 & I & -I & 0 \\
I & 0 & 0 & -I \\
-I & 0 & 0 & I \\
0 & -I & I & 0 \\
\end{pmatrix}
\end{equation}

The eigenvalues of the second operator are $(2i\Omega_{0}, -2i\Omega_{0}, 0, 0)$, with each eigenvalue corresponding to six eigenvectors. For positive frequencies, we observe a resonance at $w = 2\Omega_{0}$, since, at these frequencies, the magnitudes of the corresponding eigenvalues are minimized. Around this resonance, we can neglect the impact of other eigenvalues whose imaginary parts are not close to zero.

\vspace{10pt}

Thus, we can project the operator onto the subspace of $-2i\Omega_{0}$ eigenvectors, which do not depend on any $\Gamma$ widths or $\Omega_{0}$ and remain constant, and then invert it within that subspace. Afterward, we can translate the operator back to the initial basis. This procedure reduces the problem of 24-dimensional parametric inversion to a 6-dimensional parametric inversion. With this approach, we can finally obtain an analytical expression for the noise.

Since the expression for noise remains highly complex, we will present it only for the case $\epsilon = \Delta = 0$, with the assumptions $\delta\Gamma \ll \Gamma$, $\Gamma_{ab} \ll \Gamma_{(a, b)}$, and $L-R$ symmetry in the point contact. The first two assumptions are reasonable, since even under these conditions our result will capture the core properties of the qubit's impact on noise, while the latter assumption follows naturally from realistic physical conditions. Additionally, we will present only $\Delta S(\omega)$, the contribution of the qubit to the overall noise, as this significantly simplifies the expression and is our primary interest.

\begin{widetext}

\small

Thus, for limit $\Omega_{0} \ll \Gamma$:

\begin{equation}
    \Delta S(\omega)_{0} = \frac{2\delta\Gamma\Gamma_{a}\Gamma_{b}(\delta\Gamma_{a} + \delta\Gamma_{b})^2 (\Gamma_{a} + \Gamma_{b})(3\Gamma_{ab} + (\Gamma_{a} + \Gamma_{b}))(\Gamma_{a} + \Gamma_{a}^{'}) (\Gamma_{b} + \Gamma_{b}^{'})}{((\Gamma_{a} + \Gamma_{a}^{'})^2 + 
   (\omega - 2\Omega_{0})^2) ((\Gamma_{b} + \Gamma_{b}^{'})^2 + 
   (\omega - 2\Omega_{0})^2) ((\delta\Gamma + 
   \Gamma_{a} + \Gamma_{b} + \delta\Gamma_{a} + \delta\Gamma_{b})^2 + (\omega - 2\Omega_{0})^2) (\delta\Gamma^2 + 
   4 (\omega - 2\Omega_{0})^2)}
\end{equation}

\begin{equation}
    \Delta S(\omega)_{\Omega_{0}} = \frac{8\Omega_{0}\Gamma_{ab} \Gamma_{a} \Gamma_{b}(\delta\Gamma_{a} - \delta\Gamma_{b})^2(\Gamma_{a} + 
   \Gamma_{b})(\delta\Gamma_{a} + \delta\Gamma_{b})^2(\omega - 2\Omega_{0})}{\delta\Gamma((\Gamma_{a} + \Gamma_{a}^{'})^2 + 
   (\omega - 2\Omega_{0})^2) ((\Gamma_{b} + \Gamma_{b}^{'})^2 + 
   (\omega - 2\Omega_{0})^2) ((\delta\Gamma + 
   \Gamma_{a} + \Gamma_{b} + \delta\Gamma_{a} + \delta\Gamma_{b})^2 + (\omega - 2\Omega_{0})^2) (\delta\Gamma^2 + 
   4 (\omega - 2\Omega_{0})^2)}
\end{equation}

\normalsize

And for limit $\Gamma \ll \Omega_{0}$:

\small

\begin{equation}
    \Delta S(\omega)_{0} = \frac{2\delta\Gamma\Gamma_{a}\Gamma_{b}(\delta\Gamma_{a} + \delta\Gamma_{b})^2 (2\Gamma_{ab} + (\Gamma_{a} + \Gamma_{b}))(\Gamma_{a} + \Gamma_{a}^{'}) (\Gamma_{b} + \Gamma_{b}^{'})(\Gamma_{a} + \Gamma_{b} + \delta\Gamma_{a} + \delta\Gamma_{b})}{((\Gamma_{a} + \Gamma_{a}^{'})^2 + 
   (\omega - 2\Omega_{0})^2) ((\Gamma_{b} + \Gamma_{b}^{'})^2 + 
   (\omega - 2\Omega_{0})^2) ((\delta\Gamma + 
   \Gamma_{a} + \Gamma_{b} + \delta\Gamma_{a} + \delta\Gamma_{b})^2 + (\omega - 2\Omega_{0})^2) (\delta\Gamma^2 + 
   4 (\omega - 2\Omega_{0})^2)}
\end{equation}

\begin{equation}
    \Delta S(\omega)_{\frac{1}{\Omega_{0}}} = \frac{-16\Gamma_{ab}^2\Gamma_{a}^2\Gamma_{b}^2(\Gamma_{a} + \Gamma_{b})(\delta\Gamma_{a} + \delta\Gamma_{b})(\delta\Gamma + \delta\Gamma_{a} + \delta\Gamma_{b})(\omega - 2\Omega_{0})}{\Omega_{0}((\Gamma_{a} + \Gamma_{a}^{'})^2 + 
   (\omega - 2\Omega_{0})^2) ((\Gamma_{b} + \Gamma_{b}^{'})^2 + 
   (\omega - 2\Omega_{0})^2) ((\delta\Gamma + 
   \Gamma_{a} + \Gamma_{b} + \delta\Gamma_{a} + \delta\Gamma_{b})^2 + (\omega - 2\Omega_{0})^2) (\delta\Gamma^2 + 
   4 (\omega - 2\Omega_{0})^2)}
\end{equation}

\normalsize

\end{widetext}

Here, we denote by $\Delta S(\omega)_{0}$ the zero-th order of perturbation in terms of our limits, and the first order by $\Delta S(\omega)_{\Omega_{0}}$ or $\Delta S(\omega){\frac{1}{\Omega{0}}}$. As we can see, the zero-order correction is a product of Lorentzians, and the first-order correction introduces a small asymmetry in the total peak. Additionally, we note that $\Delta S(\omega)_{\Omega_{0}} \sim \Gamma_{ab}$, and $\Delta S(\omega){\frac{1}{\Omega{0}}} \sim \Gamma_{ab}^2$, possess opposite signs in terms of frequency space. This observation reveals the role of $\Gamma_{ab}$ — it produces an asymmetry in the noise that is not observed in the case of a simple point contact \cite{Gurvitz_2003}. Non-perturbative numerical evaluations justify this result. One can also note that, in the Zeno regime, the asymmetry is more pronounced.

\subsection{Qubit dynamics via interaction with PC 2-d "bottleneck"}

Finally, one might be interested in the reduced qubit dynamics of such a system, or the so-called "Master equation." As we discussed previously, by reducing all electron degrees of freedom, we obtain:

\begin{equation}
    \dot\sigma = (A_{q} + B_{q})\sigma
\end{equation}

However, this is a master equation for a system consisting of a qubit paired with a "2-d bottleneck" in the point contact. In the context of measurement and decoherence, we are interested only in the evolution of the qubit's density operator. Therefore, we must reduce this equation once more, by tracing over the remaining 2-d degrees of freedom in the point contact.

After partial trace, we obtain:

\begin{equation}
\begin{gathered}
    \dot\sigma_{xx} = i\Omega_{0}(\sigma_{xy} - \sigma_{yx})\\
    \dot\sigma_{xy} = (-\delta\Gamma + i\epsilon)\sigma_{xy} + i\Omega_{0}(\sigma_{xx} - \sigma_{yy}) + \delta\Gamma_{S}(\sigma_{\alpha\beta}^{xy} - \sigma_{\beta\alpha}^{xy})\\
    \dot\sigma_{yx} = (-\delta\Gamma - i\epsilon)\sigma_{yx} + i\Omega_{0}(\sigma_{yy} - \sigma_{xx}) + \delta\Gamma_{S}(\sigma_{\beta\alpha}^{yx} - \sigma_{\alpha\beta}^{yx})\\
    \dot\sigma_{yy} = i\Omega_{0}(\sigma_{yx} - \sigma_{xy})
\end{gathered}
\end{equation}

where $\delta\Gamma_{S}$ is the PC contact "symmetry width factor":

\begin{equation}
    \delta\Gamma_{S} = \frac{\delta\Gamma_{L\alpha} + \delta\Gamma_{R\alpha} - \delta\Gamma_{L\beta} - \delta\Gamma_{R\beta}}{2}
\end{equation}

\section{Generalization for any graph structures in the "bottleneck"}

Consider now any quantum dot graph structure between two contacts, depicted by fig. 8.

\begin{figure}[H]
\begin{tikzpicture}[scale=1]
    % Energy band on the left (mu_L)
    \draw[thick] (-3.5, 2) -- (-2, 2);  % Fermi energy line on the left
    \node[left] at (-3.5, 2) {$\mu_L$};  % Label for Fermi energy level on the left

    % Energy band with dashed deep blue lines for left band
    \foreach \i in {0.05, 0.15, 0.25, 0.35, 0.45, 0.55, 0.65, 0.75, 0.85, 0.95, 1.05, 1.15, 1.25, 1.35, 1.45, 1.55, 1.65, 1.75, 1.85, 1.95} {
        \draw[dashed, blue!80] (-3.5, \i) -- (-2, \i);  % Dashed lines for left energy band in deep blue
    }

    % Energy band on the right (mu_R)
    \draw[thick] (2, 0.55) -- (3.5, 0.55);  % Fermi energy line on the right
    \node[right] at (3.5, 0.55) {$\mu_R$};  % Label for Fermi energy level on the right

    % Energy band with dashed deep blue lines for right band
    \foreach \i in {0.05, 0.15, 0.25, 0.35, 0.45, 0.55} {
        \draw[dashed, blue!80] (2, \i) -- (3.5, \i);  % Dashed lines for right energy band in deep blue
    }

    % Labels for energy bands on the sides
    \node[left] at (-3.7, 1.5) {$E_L$};  % Label for left band
    \node[right] at (3.7, 1.5) {$E_R$};  % Label for right band

    % Curvilinear double-sided arrows from left band to energy levels with omega labels
    \draw[<->, thick, bend left=30] (-2.0, 1.2) to (-1.0, 1.2);  % Arrow for alpha level
    \node at (-1.5, 1.55) {$H_{TL}$};  % Omega label for alpha level arrow

    \draw[thick] (0, 1) circle (1);
    \node at (0, 1) {QG};
    
    % Curvilinear double-sided arrows from right band to energy levels with omega labels
    \draw[<->, thick, bend right=30] (2.0, 1.2) to (1.0, 1.2);  % Arrow for alpha level
    \node at (1.5, 1.5) {$H_{TR}$};  % Omega label for alpha level arrow

    % Two quantum dots for qubit representation below, positioned vertically and centered
    \draw[thick] (0, -0.5) circle (0.25);  % Left quantum dot for state 0
    \draw[thick] (0, -1.5) circle (0.25);  % Right quantum dot for state 1

    % Labels for quantum dot states (without ket notation)
    \node at (0, -0.5) {0};  % Label for the state 0
    \node at (-1, -0.5) {$+\delta\Omega$};  % Label for the state 0
    \node at (0, -1.5) {1};  % Label for the state 1

\end{tikzpicture}
\caption{Energy band diagrammatic representation of PC - arbitrary QG structure}
\end{figure}
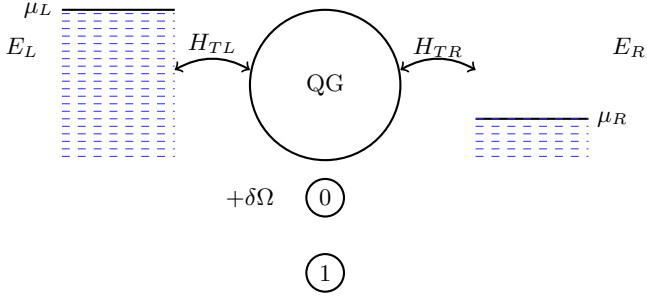

generalizing previous Hamiltonians, we can write:

\begin{equation}
\begin{gathered}
    H_{QG} = \sum_{i < j} \Omega_{ij}(a_{i}^{\dagger}a_{j} + a_{j}^{\dagger}a_{i}) + \sum_{i}E_{i}a_{i}^{\dagger}a_{i}\\
    H_{TL} = \sum_{l, i}\Omega_{li}(a_{i}^{\dagger}a_{l} + a_{l}^{\dagger}a_{i})\\
    H_{TR} = \sum_{r, i}\Omega_{ri}(a_{i}^{\dagger}a_{r} + a_{r}^{\dagger}a_{i}) 
\end{gathered}
\end{equation}

One can note that since all operators in QG are fermionic and $H_{QG}$ doesn't contain pairings like $a^{\dagger}a^{\dagger}$ or $aa$, we can diagonalize those modes by a single-particle-like Bogolyubov transformation and obtain a new representation of $H_{QG}$:

\begin{equation}
    H_{QG} = \sum_{i}E'_{i}b_{i}^{\dagger}b_{i}
\end{equation}

and:

\begin{equation}
    a_{i} = \sum_{j}U_{ij}b_{j}
\end{equation}

where $U_{ij}$ is an unitary $n$ dimensional operator and $n$ is the number of dots in $QG$. The same linear transformation will, obviously, hold for any transition amplitudes $\Omega_{ij}$ in the new fermionic "island" basis:

\begin{equation}
    \Omega^{'}_{lk} = \Omega_{li}U_{ik}
\end{equation}

\vspace{10pt}

Now, we can rewrite our Hamiltonian in terms of the new creation-annihilation operators, and our system will begin to resemble the previous one, which we analyzed in detail before, but for the case of $n$ independent energy levels on the "island", fig. 9:

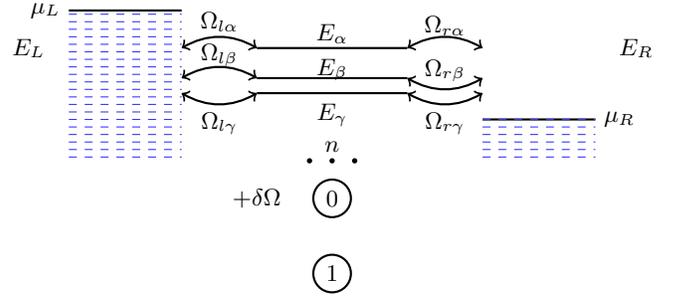
\begin{figure}[H]
\begin{tikzpicture}[scale=1]
    % Energy band on the left (mu_L)
    \draw[thick] (-3.5, 2) -- (-2, 2);  % Fermi energy line on the left
    \node[left] at (-3.5, 2) {$\mu_L$};  % Label for Fermi energy level on the left

    % Energy band with dashed deep blue lines for left band
    \foreach \i in {0.05, 0.15, 0.25, 0.35, 0.45, 0.55, 0.65, 0.75, 0.85, 0.95, 1.05, 1.15, 1.25, 1.35, 1.45, 1.55, 1.65, 1.75, 1.85, 1.95} {
        \draw[dashed, blue!80] (-3.5, \i) -- (-2, \i);  % Dashed lines for left energy band in deep blue
    }

    % Energy band on the right (mu_R)
    \draw[thick] (2, 0.55) -- (3.5, 0.55);  % Fermi energy line on the right
    \node[right] at (3.5, 0.55) {$\mu_R$};  % Label for Fermi energy level on the right

    % Energy band with dashed deep blue lines for right band
    \foreach \i in {0.05, 0.15, 0.25, 0.35, 0.45, 0.55} {
        \draw[dashed, blue!80] (2, \i) -- (3.5, \i);  % Dashed lines for right energy band in deep blue
    }

    % Labels for energy bands on the sides
    \node[left] at (-3.7, 1.5) {$E_L$};  % Label for left band
    \node[right] at (3.7, 1.5) {$E_R$};  % Label for right band

    % Two energy levels in the middle, closer and lower in the band
    \draw[thick] (-1, 1.5) -- (1, 1.5);  % Alpha level, placed lower in the band

    \draw[thick] (-1, 0.9) -- (1, 0.9);  % Beta level, placed lower and closer to alpha

    \draw[thick] (-1, 1.1) -- (1, 1.1);  % Beta level, placed lower and closer to alpha

    % Labels for energy levels E_alpha and E_beta, adjusted for clarity
    \node[above] at (0, 1.45) {$E_\alpha$};  % Label for alpha energy level
    \node[above] at (0, 0.95) {$E_\beta$};   % Label for beta energy level
    \node[above] at (0, 0.35) {$E_\gamma$};   % Label for beta energy level

    % Curvilinear double-sided arrows from left band to energy levels with omega labels
    \draw[<->, thick, bend left=30] (-2.0, 1.5) to (-1.0, 1.5);  % Arrow for alpha level
    \node at (-1.5, 1.85) {$\Omega_{l\alpha}$};  % Omega label for alpha level arrow

    \draw[<->, thick, bend left=30] (-2.0, 1.1) to (-1.0, 1.1);  % Arrow for alpha level
    \node at (-1.5, 1.4) {$\Omega_{l\beta}$};  % Omega label for alpha level arrow

    \draw[<->, thick, bend left=30] (2.0, 1.1) to (1.0, 1.1);  % Arrow for alpha level

    \draw[<->, thick, bend left=-30] (-2.0, 0.9) to (-1.0, 0.9);   % Arrow for beta level
    \node at (-1.5, 0.5) {$\Omega_{l\gamma}$};   % Omega label for beta level arrow

    % Curvilinear double-sided arrows from right band to energy levels with omega labels
    \draw[<->, thick, bend right=30] (2.0, 1.5) to (1.0, 1.5);  % Arrow for alpha level
    \node at (1.5, 1.8) {$\Omega_{r\alpha}$};  % Omega label for alpha level arrow

    \node at (1.5, 1.2) {$\Omega_{r\beta}$};  % Omega label for alpha level arrow

    \draw[<->, thick, bend right=-30] (2.0, 0.9) to (1.0, 0.9);   % Arrow for beta level
    \node at (1.5, 0.5) {$\Omega_{r\gamma}$};   % Omega label for beta level arrow

    % Two quantum dots for qubit representation below, positioned vertically and centered
    \draw[thick] (0, -0.5) circle (0.25);  % Left quantum dot for state 0
    \draw[thick] (0, -1.5) circle (0.25);  % Right quantum dot for state 1

    % Labels for quantum dot states (without ket notation)
    \node at (0, -0.5) {0};  % Label for the state 0
    \node at (-1, -0.5) {$+\delta\Omega$};  % Label for the state 0
    \node at (0, -1.5) {1};  % Label for the state 1

    \fill (-0.3,0) circle (1pt);
    \fill (0.0,0) circle (1pt);
    \fill (0.3,0) circle (1pt);
     \node[above] at (0, 0) {$n$};
\end{tikzpicture}
\caption{Generalized diagrammatic representation of an PC - one electron transistor like energy band}
\end{figure}

Revisiting the previous derivations, we can establish universal rules for obtaining Master Equations in any "island" system:

\begin{equation}
\hspace{-0.5cm}
\begin{gathered}
    \dot\sigma_{(\alpha\beta\gamma...)(\alpha'\beta'\gamma'...)}^{n} = \sum_{a, b}(-1)^{(a - b)}\Gamma_{L(a,b)}\sigma_{(\alpha\beta\gamma...\overline{a}...)(\alpha'\beta'\gamma'...\overline{b}...)}^{n}\\
    + \sum_{a, b}\Gamma_{R(a,b)}\sigma_{(\alpha\beta\gamma...a...)(\alpha'\beta'\gamma'...b...)}^{n - 1}\\
     + \sum_{a, b}\frac{\Delta\Gamma_{ab}}{2}\sigma_{(\alpha\beta\gamma...a\overline{b}...)(\alpha'\beta'\gamma'...)}^{n} + \sum_{a, b}\frac{\Delta\Gamma_{ab}}{2}\sigma_{(\alpha\beta\gamma...)(\alpha'\beta'\gamma'...a\overline{b}...)}^{n}\\
     -(\sum_{a}\frac{\Gamma_{La}}{2} + \sum_{b}\frac{\Gamma_{Rb}}{2} + i(\sum_{\alpha, \beta, \gamma...}E_{k} - \sum_{\alpha', \beta', \gamma'...}E_{l}))\sigma_{(\alpha\beta\gamma...)(\alpha'\beta'\gamma'...)}^{n}
\end{gathered}
\end{equation}

The first sum represents all terms that differ from the left one by the insertion of two indexes in the bra and ket parts, the second sum represents all terms that differ from the left one by the removal of two indexes in the bra and ket parts, and the last two sums account for the change of index in the bra and ket parts, respectively. In the diagonal part, for left widths, we sum over all indexes that are absent in the ket and bra, and for right widths, we sum over all indexes which are present in the ket and bra. Obtaining the dynamics, including the qubit, is therefore trivial: one should apply "variation" to all coefficients in the equations, as was done for the 2-d case, and flip the coefficients for coherent terms.

Lastly, we can analogously derive qubit's master equation in case of arbitrary QG. It will still keep the past form, but with generalized $\delta\Gamma_{S}$:

\begin{equation}
    \delta\Gamma_{S(\alpha\beta\dots, \gamma\zeta \dots)} = \sum_{z = \alpha, \beta \dots} \frac{\delta\Gamma_{Lz} + \delta\Gamma_{Rz}}{2} - \sum_{z = \gamma, \zeta \dots} \frac{\delta\Gamma_{Lz} + \delta\Gamma_{Rz}}{2}
\end{equation}

Using this approach, one can calculate the current and it's noise for more complex structures right out of the box. We will show only one simple example.

\hspace{10pt}
Imagine that we have a circular quantum dot graph, a structure often examined in the context of quantum dot evaluations and their models \cite{Solenov_2006}, depicted by the fig. 10.
\hspace{-10pt}

\begin{figure}
\begin{tikzpicture}[transform canvas={xshift=0cm, yshift = -2cm}]
    \draw[thick] (0, 1) circle (1);
    \node at (0, 1) {QG};
    \node at (0, -1) {\Huge \rotatebox{90}{$=$}};
\end{tikzpicture}

\begin{tikzpicture}[scale=1, 
  dot/.style={circle, draw, fill=blue!30, inner sep=2.5pt},
  qubit/.style={circle, draw, fill=red!30, inner sep=3pt},
  arrow/.style={<->, thick}, transform canvas={xshift=0cm, yshift = -5cm}]

% Parameters
\def\R{3} % Radius of the ring
\def\Qdist{4} % Distance of qubit below the ring
\def\N{12} % Number of quantum dots in the ring

% Draw the ring of quantum dots
\foreach \i in {1,...,\N} {
    % Compute angle for each dot
    \pgfmathsetmacro{\angle}{360/\N * (\i - 1)}
    \pgfmathsetmacro{\x}{\R * cos(\angle)}
    \pgfmathsetmacro{\y}{\R * sin(\angle) * 0.5}
    \node[dot] (QD\i) at (\x, \y) {};
}

% Draw tunneling amplitudes (bidirectional arrows between neighboring dots)
\foreach \i in {1,...,\N} {
    \pgfmathsetmacro{\next}{mod(\i, \N) + 1}
    \draw[arrow] (QD\i) -- (QD\next) node[midway, above, sloped] {\(\Omega\)};
}

\end{tikzpicture}
\vspace{7cm}
\caption{Schematic representation of a cyclic quantum dot graph structure, which will participate in PC quantum transport.}
\end{figure}
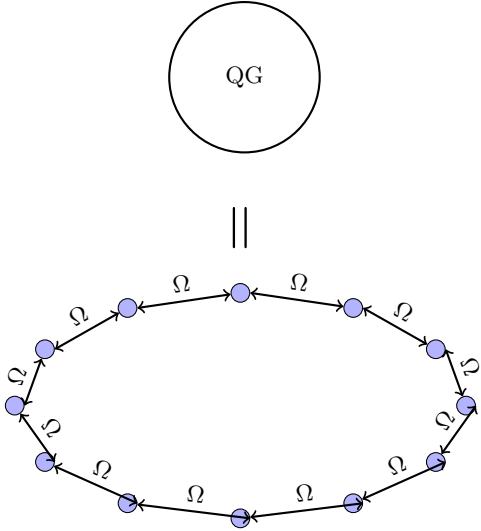

\vspace{0pt}

It's a known fact that diagonalization of such structure will result in:

\begin{equation}
    a_{l} = \sum_{j}\frac{e^{i\frac{2\pi lj}{N}}b_{j}}{\sqrt{N}}
\end{equation}

So, we can conclude, that such structure will result in particularly interesting outcome:

\begin{equation}
\begin{gathered}
    \Gamma_{nk} = \frac{\Gamma}{N}\cos(\frac{2\pi(n - k)}{N})\\
    E_{n} = 2\Omega\cos(\frac{2\pi n}{N})
\end{gathered}
\end{equation}

where $n, k$ denote indexes of new fermionic states, describing degrees of freedom in $QG$. One can check that, in the limit of low tunneling in the ring, for stationary current, we get from the obtained equations:

\begin{equation}
    I(\infty) \sim \frac{\Omega^2}{\Gamma} 
\end{equation}

which totally coincides with natural absence of current without tunneling in the ring.

Finally, one can obtain operators $A, B, A_{q}, B_{q}$, using rules, described before.

\section{Including relaxation in non-unitary dynamics}

Up to this point, we have excluded relaxation from our consideration.
One can formally include it by starting directly from the system Hamiltonian and accounting for the interaction of the qubit with an arbitrary boson bath in the Lee model, as done in \cite{Gurvitz_2003}. Nevertheless we can still capture the key effects of relaxation on our results.

Based on the results of \cite{Gurvitz_2003}, we can conclude that relaxation effectively modifies all $\delta \Gamma$ spectral widths related to the qubit:

\begin{equation}
    \delta \Gamma \rightarrow \delta \Gamma + \Gamma_{r}
\end{equation}

where $\Gamma_{r}$ is the relaxation width, which can be obtained using the same many-body wave function method. This shift will "diffuse" the qubit-related peaks in the noise spectrum. Common factor $\delta\Gamma$ in $\Delta S(\omega)_{0}$ is untouched, because it is proportional to the current difference due to qubit Couloumb interaction with PC.

The simplest approach to incorporate relaxation into the qubit Master equation gives us, in the case of $\epsilon = 0$:

\begin{equation}
\begin{gathered}
    \dot\sigma_{xx} = (\dots) - \frac{\Gamma_{r}}{2}(\sigma_{xx} - \frac{1}{2})\\
    \dot\sigma_{xy} = (\dots) - \frac{\Gamma_{r}}{2}\sigma_{xy}
\end{gathered}
\end{equation}

Finally, we note an indication of the fundamental role of relaxation, which follows from the properties of the noise spectrum. One can see that if we set $\Gamma_{r} = 0$ and take the limit $\delta \Gamma \rightarrow 0$, which in our case represents an infinitely small coupling of the qubit to the PC, $\Delta S(2\Omega_{0}) \rightarrow C$ with $|C| > 0$. This implies that the qubit's impact on the PC current noise statistics always has a finite value, which can also be verified numerically and, thus, is not an artifact of approximations.

One might expect the opposite result, as setting $\delta\Omega = 0$ (which was analyzed earlier) makes the PC and qubit independent of each other, strictly leading to $\Delta S(\omega) = 0$. This non-physical result disappears only when finite relaxation rates are introduced: with $\Gamma_{r} > 0$, we effectively obtain $\Delta S(\omega) \rightarrow 0$. Thus, finite relaxation is necessary for physically consistent current statistics. These considerations are also applicable to the results of \cite{Gurvitz_2003}.

\section{Conclusions}

We formalized the many-body wave-function approach for deriving Master equations in quantum transport systems including systems interacting with qubit. Using these methods, we obtained the tunnel current noise power spectrum, up to first-order corrections for both key $\Omega_{0}/\delta\Omega$ limits, and the reduced master equation describing the qubit's degrees of freedom. In the coherent regime, we acquired an analytical asymmetry in the noise spectra for analogous PC qubit measurement apparatus which has not been obtained before. The criteria for the enclosed qubit's Master equation, excluding PC degrees of freedom, are derived.

Using these results, we generalized the derivations for arbitrary quantum graph systems in PC, reducing their dynamics to "one electron-transistor" structure via diagonalization. Systems considered are shown to be capable of producing effective coherent widths of the same order as ordinary level widths.

\bibliography{references}

\end{document}